\documentclass[twocolumn,showpacs,preprintnumbers,amsmath,amssymb]{revtex4-1}

\usepackage{epsfig}
\usepackage{dcolumn}
\usepackage{bm}
\usepackage[dvipsnames]{xcolor}
\usepackage{ulem}

\begin{document}

\preprint{\today} 

\title{Technique for a direct measurement of the cesium anapole moment using coherent rf and Raman interactions}

\author{Amy Damitz$^{1,2}$, Jonah A. Quirk$^{1,2}$, Carol E. Tanner$^3$, and D. S. Elliott$^{1,2,4}$}

\affiliation{%
   $^1$Department of Physics and Astronomy, Purdue University, West Lafayette, Indiana 47907, USA\\
   $^2$Purdue Quantum Science and Engineering Institute, Purdue University, West Lafayette, Indiana 47907, USA\\
   $^3$Department of Physics and Astronomy, University of Notre Dame, Notre Dame, Indiana 46556, USA\\
    $^4$The Elmore Family School of Electrical and Computer Engineering, Purdue University, West Lafayette, Indiana 47907, USA
   }

\date{\today}

\begin{abstract}

 We report progress toward measurements of the electric dipole (E1) transition moments between hyperfine components of the ground state of atomic cesium. This transition is weakly E1 allowed due to weak interactions between nucleons within the nucleus, which lead to a parity-odd current distribution and its associated anapole moment.  
In this report, we discuss the experimental geometry of our measurement scheme, explore the effects of extraneous fields that can obscure the signal, present initial measurements, analyze the sources and magnitudes of measurement noise, and suggest improvements to the current apparatus.  
\end{abstract}

\maketitle 

\section{Introduction}\label{sec:introduction}
Parity nonconserving (PNC) weak force interactions between nucleons can lead to a parity-odd, time-reversal-conserving current distribution within the nucleus, generating an anapole moment~\cite{Zeldovich1958}. 
In 1984, Flambaum, Khriplovich, and Sushkov~\cite{FlambaumKS1984} derived an approximate expression for the nuclear anapole moment, and determined a dimensionless anapole moment coefficient $\kappa_a$ for $^{133}$Cs, $^{203,205}$Tl, $^{209}$Bi, and $^{207}$Pb. 
The only significant observation of an anapole moment in any atomic species was carried out in atomic cesium by the Boulder group in 1997~\cite{WoodBCMRTW97}.  In that work, the investigators measured the PNC moments of the $6s \ ^2S_{1/2} F \rightarrow 7s \ ^2S_{1/2} F^{\prime}$ transitions, where $F$ and $F^{\prime}$= 3 or 4 are the total angular momenta of the atom (electronic $J$ plus nuclear spin $I$) in the initial and final states respectively. Haxton and Wieman~\cite{HaxtonW01} later extracted the anapole moment from the nuclear-spin-dependent (NSD) component of the PNC moments, i.e.\ the difference between the $F = 3 \rightarrow F^{\prime} = 4$ and the $F = 4 \rightarrow F^{\prime} = 3$ transitions. However, this result is in conflict with meson coupling constants derived from a number of nuclear scattering experiments, as listed in Ref.~\cite{HaxtonW01}, and there has long been a need for a new determination of the anapole moment of any atomic species as a means of either confirming the Boulder results or providing a new value. (See Ref.~\cite{SafronovaBDKDC2018} for a comprehensive review of atomic parity violation and other searches for new physics in atoms.)  We report progress on such a measurement in the present paper.

The weak force interaction between constituents of an atom, mediated through exchange of the $Z^0$ boson, can be observed through precision measurements of transitions forbidden by selection rules for normal (i.e.\ electric dipole (E1), magnetic dipole (M1), electric quadrupole (E2), etc.)\ optical transitions. The dominant PNC interaction is typically nuclear spin independent (NSI) and is a result of the nucleon vector current coupling to the electron axial-vector current ($V_n A_e$). A precision measurement of this interaction produces a value of the weak charge $Q_w$ of the nucleus. For example in cesium, the Boulder group's measurement~\cite{WoodBCMRTW97} led to $Q_w = -73.66 \: (34)$ (using the atomic structure calculation result of Ref.~\cite{PorsevBD09,PorsevBD10}) or $Q_w = -73.07 \: (43)$ (using the atomic structure calculation of Ref.~\cite{DzubaF12}), as reported in Ref.~\cite{TohDTJE2019}. This is the most precise NSI measurement of this type, and is in agreement with the standard model value $Q_w^{sm} = -73.23 \: (1)$~\cite{RPP2018}.   

Alternatively, the NSD contribution to the PNC amplitude is due to three different effects: the nuclear anapole moment, as introduced above; the axial-vector nucleon and vector electron currents ($A_n V_e$); and the combination of hyperfine mixing and the ($V_n A_e$) interaction current. These interactions can be written in the form~\cite{JohnsonSS03}
  \begin{equation}\label{eq:NSDHamiltonian}
    H^{(i)} = \frac{G}{\sqrt{2}} \kappa_i \boldsymbol{\alpha} \cdot \mathbf{I} \rho(r), 
  \end{equation}
where $G \simeq 1.166 \times 10^{-5}$ GeV$^{-2}$ is the Fermi constant, $\boldsymbol{\alpha}$ is the Dirac spin matrix, $\mathbf{I}$ is the nuclear spin, and $\rho(r)$ is the nuclear density. 
The subscript $i$, where $i \in \{a, 2, {\rm hf}\},$ indicates the interaction type. 
$a$ is the anapole moment, 2 is the ($A_n V_e$) currents, and hf is the combined hyperfine interaction with the ($V_n A_e$) current. The $\kappa_i$ coefficients characterize the strength of these NSD interactions.  It was also shown in Ref.~\cite{JohnsonSS03} that, while Eq.~(\ref{eq:NSDHamiltonian}) is not strictly correct for the combined hyperfine ($V_n A_e$) contribution, their more rigorous calculations showed only a small difference, and they derived an effective coefficient $\kappa_{\rm hf}$ for this interaction. Thus the total effective NSD coefficient is 
\begin{equation}\label{eq:kappa}
    \kappa = \kappa_a + \kappa_2 + \kappa_{\rm hf}.
\end{equation} 
Johnson, Safronova, and Safronova~\cite{JohnsonSS03} also calculated $\kappa_{\rm hf}$, including correlation corrections, and found $\kappa_{\rm hf} = 0.0049$, a value that was 40\% smaller than that calculated previously~\cite{BouchiatP91}. When $\kappa_2 = 0.0140$~\cite{JohnsonSS03} and $\kappa_{\rm hf}$ are subtracted from the total spin-dependent moment $\kappa = 0.117 \; (16)$ determined from the measurement by Wood et al.~\cite{WoodBCMRTW97}, this leads to a larger value for $\kappa_a = 0.098 \; (16)$, increasing the discrepancy between this measurement and the value obtained from meson coupling constants~\cite{JohnsonSS03,HaxtonW01,desplanques1980}.

Johnson, Safronova, and Safronova~\cite{JohnsonSS03} also calculated the PNC electric dipole transition moment
\begin{equation}\label{eq:epnc}
    \mathcal{E}_{\rm PNC} = i  \langle 6s_{1/2} \  F^{\prime}=4||er||6s_{1/2} \ F=3 \rangle 
\end{equation}
 between the hyperfine components of the $6s _{1/2}$ state in atomic cesium, $ \mathcal{E}_{\rm PNC} =  i \kappa \ 1.724 \times 10^{-11} e a_0$, where $e$ is the elementary charge and $a_0$ is the Bohr radius. (We use the short-hand spectroscopic $n \ell_j$ notation in place of $n \ell \; ^2L_J$.) In a more recent work~\cite{DzubaF12}, Dzuba and Flambaum included Brueckner-type correlations, increasing this quantity by 6\%, $ \mathcal{E}_{\rm PNC} =  i \kappa \ 1.82 \times 10^{-11} e a_0$.

Measurements (or plans for measurements) of nuclear anapole moments in various other atomic or molecular species have been reported in several earlier works~\cite{VetterMMLF1995,Bouchiat2007,GomezASOD2007,AltuntasACD2018}.  The Fortson group reported~\cite{VetterMMLF1995} a $1 \sigma$ determination of the anapole moment of thallium derived from the NSD parity violation amplitude $\mathcal{E}_{\rm PNC}$ for an optical transition in that heavy element. Bouchiat~\cite{Bouchiat2007} considered the effect of the anapole moment in cesium, and showed theoretically that this effect would lead to a linear dc Stark shift of the transition frequency between hyperfine components of the ground state, which she calculated as $\sim 7 \: \mu$Hz for a dc electric field strength of 100 kV/cm and an optical power of 1 kW. Investigations of the anapole moment of the francium nucleus were proposed in Ref.~\cite{GomezASOD2007}. Francium has no stable isotopes, necessitating studies at facilities such as the Isotope Separator and Accelerator (ISAC) at TRIUMF in Vancouver, Canada. 
These investigators plan to use a technique similar in some respects to that used in our work; they plan to prepare the atoms in a superposition state using an optical Raman interaction, and measure $ \mathcal{E}_{\rm PNC}$ for an E1 transition between hyperfine components of the ground state using a microwave field (frequency $\sim$45 GHz, with a wavelength of 0.7 cm).  Several isotopes of francium are produced at their facility, allowing studies as a function of neutron number $N$. They plan to minimize M1 interactions by confining the microwave field within an open microwave Fabry-Perot resonator, and trapping the atoms at a  magnetic field node of the standing wave pattern. The anapole moment of francium is expected to be larger than that in cesium by a factor of $\sim$11. 
Finally, NSD parity violation measurements have been pursued in BaF~\cite{AltuntasACD2018}. In this molecule, the investigators take advantage of the small energy spacing between rotational states of the molecule, which they Zeeman tune into degeneracy. They observe population transfer  between molecular states when the molecule is driven by a time-dependent electric field, using Stark interference as a means of calibrating the measurement.
While even isotopes of Ba such as $^{138}$Ba, which is the isotope of their measurements completed to this point, have nuclear spin $I=0$, and therefore do not have an anapole moment, this work serves as a check on the sensitivity of a future measurement with a Ba isotope that does have nuclear spin and an anapole moment, such as $^{137}$Ba. They report a measurement sensitivity about 10 times smaller than the expected signal size.

We have been working toward a measurement of $ \mathcal{E}_{\rm PNC}$ in atomic cesium. In the present work, we discuss our progress toward this goal.
The measurement is carried out in an atomic beam system, in which we prepare the atoms in a 50:50 superposition state before interacting with an intense 9.2 GHz electric field confined to a cylindrical radio frequency (rf) cavity.  
In section \ref{sec:interactions} of the paper we discuss the various interactions of the rf field with the ground state atoms that are possible in our setup. We then explore in detail how we control these interactions in an rf cavity in section \ref{sec:rfcavity}, and the experimental geometry used in section IV. Next up are the experimental results that we have obtained to date, and the likely source of signal that currently obscures the signature of the anapole contribution in sections \ref{sec:popmod} and \ref{sec:ModSpec}. Finally we discuss future plans, which we expect will allow a successful measurement of the anapole moment, in section \ref{sec:four_input_cavity} and conclude in section \ref{sec:Conclusion}.

\section{rf Interactions}\label{sec:interactions} 

\begin{figure}[b!]
 \begin{centering}
 	  \includegraphics[width=0.35\textwidth]{
    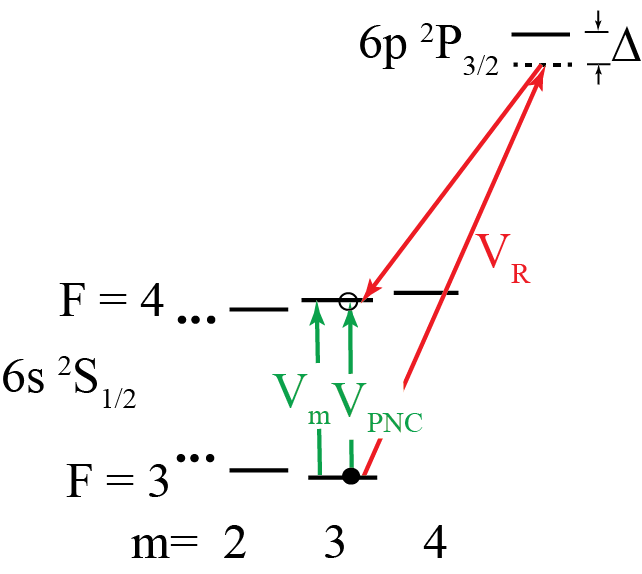}\\

  \caption{ Partial energy level diagram and the interaction transitions (not to scale). The Raman transition $V_R$ is used to prepare the atoms in a superposition state before they enter the rf cavity.  The strong magnetic dipole transition $V_m$ must be suppressed to allow the measurement of $V_{PNC}$. $\Delta$ is the detuning of the Raman lasers from direct excitation of the $6p_{3/2}$ state.}
        \label{fig:energylevels}
 \end{centering}
 \end{figure}

Our goal is to measure $ \mathcal{E}_{\rm PNC}$ as defined in Eq.~(\ref{eq:epnc}).  Fig.~\ref{fig:energylevels} is a partial energy level diagram showing the low lying energy levels of cesium and the relevant interactions. 
The ground state E1 transition is weakly allowed due to perturbations caused by the weak force between the quarks of the nucleus. A measurement of $ \mathcal{E}_{\rm PNC}$  will allow us to determine the coefficient $\kappa$ of Eq.~(\ref{eq:kappa}), which quantifies the strength of the weak interaction. The specific transitions chosen for these measurements are $\psi_g = 6s_{1/2} \: F=3, m \rightarrow \psi_e = 6s_{1/2} \: F^{\prime} = 4, m$, where $m = 3$ or $-3$ is the magnetic sublevel of the state of total angular momentum $F$.
 The primary challenge in this measurement is to reduce other electromagnetic field components that could drive competing parity conserving interactions between the same initial and final states. 

\begin{figure*}
 \begin{centering}
 	  \includegraphics[width=0.15\textwidth]{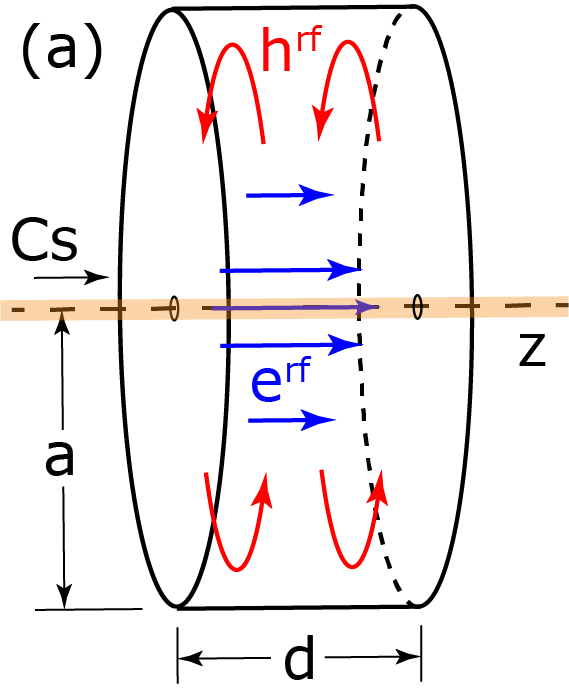} \hspace{0.2in}
    \includegraphics[width=0.31\textwidth]{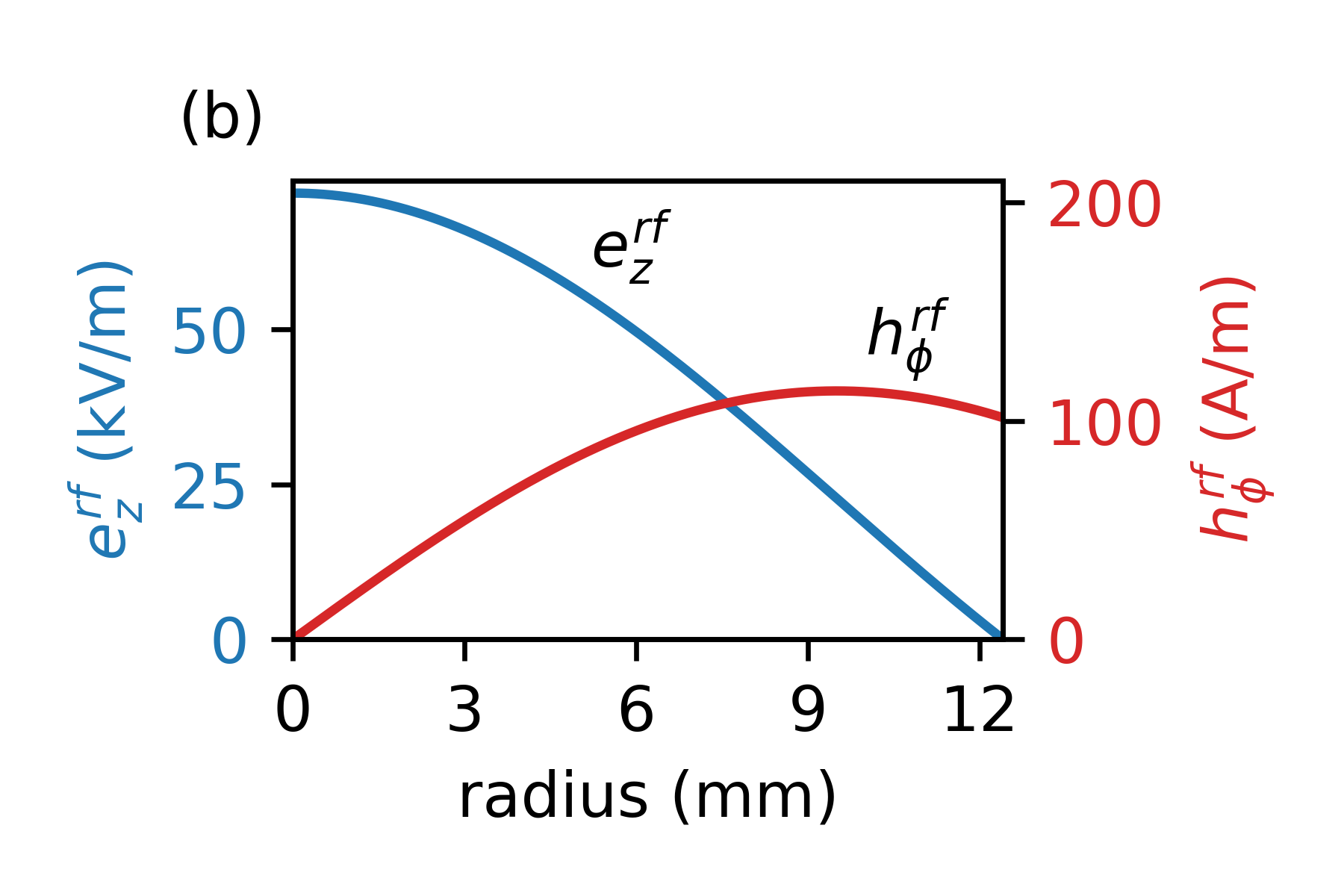}  \hspace{0.2in}
       \includegraphics[width=0.3\textwidth]{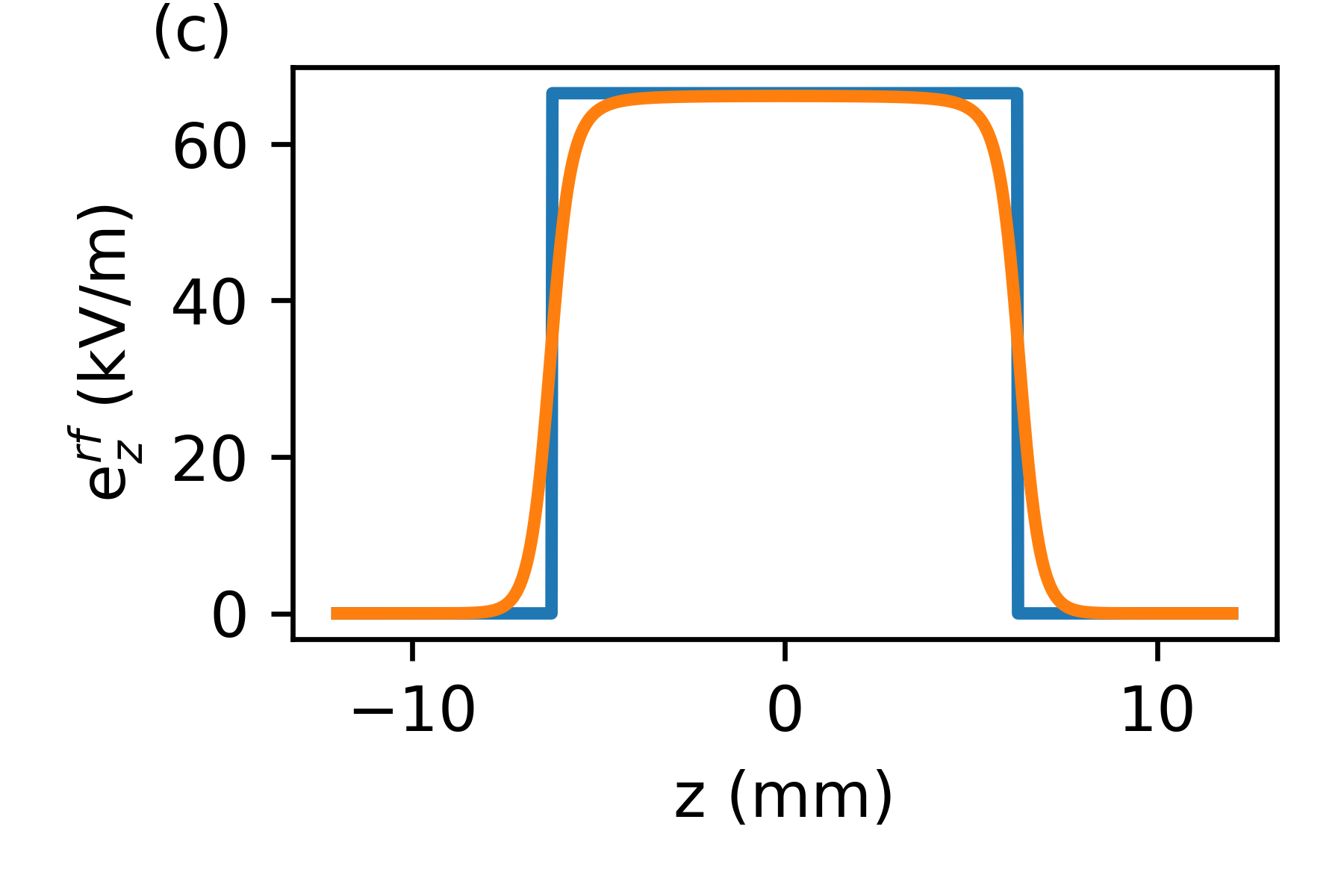}\\
  \caption{(a) Simplified diagram of the cylindrical rf cavity, showing the axial electric field $\mathbf{e}^{\rm rf}$ and the azimuthal magnetic field $\mathbf{h}^{\rm rf}$ of the TM$_{010}$ mode.  The dimensions of the cavity are $a = 1.241$ cm and $d = 1.25$ cm. (b) Plots of $e_z^{\rm rf}$ and $h_{\varphi}^{\rm rf}$ vs.\ the radial distance $\rho$. (c)  Results of numerical simulations (orange) of $e_z^{\rm rf}(z)$ vs. $z$ (along the atom beam) compared to $e_z^{\rm rf}(z)$ for a perfect cavity with no holes (blue).}
        \label{fig:Electricfieldrod}
 \end{centering}
 \end{figure*}

We use the electric field $\mathbf{e}^{\rm rf}$ of the TM$_{010}$ mode of a cylindrical rf cavity, as shown in Fig.~\ref{fig:Electricfieldrod}(a).
(In this work, lower case characters represent oscillating field amplitudes, while upper case characters are used for static fields.)  This field $\mathbf{e}^{\rm rf} = e_z^{\rm rf}(\rho) \hat{z}$ is directed in the axial direction (which we label the $z$-axis), with magnitude $e_z^{\rm rf}(\rho) = e_0^{\rm rf} J_0(p_{0,1} \rho/a)$. $e_0^{\rm rf}$ is the field amplitude on the axis of the cavity, $\rho$ is the radial distance from the axis, $p_{0,1}=2.4048$ is the first zero of the zeroth-order Bessel function, and $a = 1.241$ cm is the radius of the cavity. A plot of this field amplitude vs.\ the radius $\rho$ is shown in Fig.~\ref{fig:Electricfieldrod}(b) as the blue line.
The atoms pass through the cavity along the $z$-axis, and experience a relatively uniform field magnitude while inside the cavity. The field amplitude vs.\ $z$ is shown in Fig.~\ref{fig:Electricfieldrod}(c), both for an ideal cavity with no holes or perturbations (blue line), and for our specific cavity geometry (orange line), as computed numerically (COMSOL Multiphysics).  The length of the cavity is $d=1.25$ cm.
The rf magnetic field of this mode is azimuthal, $\mathbf{h}^{\rm rf} = h_{\phi}^{\rm rf}(\rho) \hat{\phi}$, where $h_{\phi}^{\rm rf}(\rho) = ( i \omega \varepsilon_0 a / p_{0,1}) e_0^{\rm rf}  J_1(p_{0,1} \rho/a)$, and $J_1$ is the first-order Bessel function.  We show plots of $e_z^{\rm rf}(\rho)$ and $h_{\phi}^{\rm rf}(\rho)$ vs.\ $\rho$ in Fig.~\ref{fig:Electricfieldrod}(b).  Note that at the location of the atomic beam ($\rho \ll a$), the electric field of this cavity mode is near its maximum value, while the magnetic field is minimal.

We treat the atom as a two-level system, where the state of the atoms is
   \begin{equation}\label{eq:superposition}
     \psi=c_g(t)\psi_ge^{-i\omega_gt}+c_e(t)\psi_ee^{-i\omega_et}.
\end{equation}
$c_g(t)$ ($c_e(t)$) is the time-varying state amplitude of the ground (excited) state.  The state amplitudes change due to interactions $V_{m}$ and $V_{\rm PNC}$ (defined explicitly below) with the applied rf fields in the cavity, or $V_R$ with the pair of Raman laser fields which intersect the atom beam before the rf cavity. $V_{\rm R}$ prepares the system in a 50:50 mixture of $\psi_g$ and $\psi_e$. 
The equations of motion of the amplitudes $c_g(t)$ and $c_e(t)$, as derived from the Schr\"{o}dinger Equation, are~\cite{meystre2007elements,foot2005atomic} 
\begin{equation}
    \frac{dc_e(t)}{dt}=c_g(t)\left(-\frac{i}{\hbar}V e^{i\Delta \omega t}\right)
    \label{eq:equationofmotionexcited}
\end{equation}
and
\begin{equation}
    \frac{dc_g(t)}{dt}=c_e(t)\left(-\frac{i}{\hbar}(V)^*e^{-i\Delta \omega t}\right),
    \label{eq:equationofmotionground}
\end{equation}
where
\begin{equation}
  V=V_{R}+V_{m}+V_{\rm PNC}
\end{equation}
and $\Delta \omega =\omega_{e}-\omega_{g}$.

The interaction potential for the E1 transition between hyperfine components of the ground state is
\begin{equation}\label{eq:V_PNC}
   V_{\rm PNC} = - \boldsymbol{\mathcal{E}}_{\rm PNC} \cdot \mathbf{e}^{\rm rf}.
\end{equation}
Since $\mathbf{e}^{\rm rf}$ is oriented along the $z$-axis parallel to a dc magnetic field, this interaction drives a $\Delta m = 0$ transition.  
We determine the strength of the interaction by measuring the change in population $|c_e(t)|^2$ of the ground state $F=4$ hyperfine level.  
This interaction is quite small, however, and in the presence of other interactions (primarily magnetic dipole interactions, as shown in Fig.~\ref{fig:energylevels}), its measurement can be extremely difficult. In this section, we describe the other interactions that must be minimized if the PNC term is to be successfully measured.  

The first unwanted interaction that we consider is the magnetic dipole interaction,
\begin{equation}
   V_m = - \boldsymbol{\mu} \cdot \mathbf{b}^{\rm rf} , 
\end{equation}
where $\mathbf{b}^{\rm rf}$ is the magnetic flux density of the rf mode.
The magnetic moment $\boldsymbol{\mu}$ in terms of orbital, electron spin, and nuclear angular momenta, $\mathbf{L}$, $\mathbf{S}$, and $\mathbf{I}$, respectively, is 
\begin{equation}
    \boldsymbol{\mu} = -\mu_B \left( \mathbf{L} + g_s \mathbf{S} \right) + g_I \mu_N \mathbf{I},
\end{equation}
where $\mu_B = 9.274 \times 10^{-24}$ J/T is the Bohr magneton and $\mu_N = 5.051 \times 10^{-27}$ J/T is the nuclear magneton.  $g_s \approx 2$ ($g_I \approx 2.582$) is the electron (nuclear) gyromagnetic ratio. $\mathbf{L} = 0$ for the ground $S$ state, the spin angular momentum operator is $\mathbf{S} = (\hbar/2) \boldsymbol{\sigma}$, where $\boldsymbol{\sigma}$ are the Pauli spin matrices, and $\mu_N \ll \mu_B$, allowing us to ignore the nuclear spin term.  Therefore the magnetic interaction can be written as 
\begin{equation}\label{eq:V_m1}
   V_m = \frac{1}{2} g_s \mu_B \mu_0 \: \boldsymbol{\sigma} \cdot \mathbf{h}^{\rm rf}.  
\end{equation}
$\mathbf{h}^{\rm rf} = \mathbf{b}^{\rm rf}/\mu_0$ is the rf magnetic field and $\mu_0$ is the magnetic constant. The immediate inference of Eq.~(\ref{eq:V_m1}) then, is that the rf magnetic field component $h_z^{\rm rf}$ must be minimized, as this component drives a $\Delta m = 0$ transition that could easily mask the electric dipole signal that we wish to measure. One important benefit of confining the rf field to the TM$_{010}$ mode of a cylindrical cavity, and propagating the atoms along the axis of this cavity, is that $h_z^{\rm rf}$ is highly suppressed. We note that the holes in the final machined rf cavity, which are necessary for transmission of the rf power or atoms, can potentially induce $h_z^{\rm rf}$, which we will discuss in Sections \ref{sec:ModSpec} and \ref{sec:four_input_cavity}.

We apply a modest ($\sim7$ G) static magnetic flux density $\mathbf{B}$ to the interaction region to Zeeman split the magnetic substates of both hyperfine components of the ground state and to adjust the transition frequency of the $\psi_g \rightarrow \psi_e$ line to the resonance of the rf cavity.
If $\mathbf{B}$ is not perfectly aligned with the axis of the cavity, then the transverse components of the magnetic field $B_x$ and $B_y$ can mix magnetic substates.  
As a result, nominally $\Delta m = \pm1$ transitions driven by $h_x^{\rm rf}$ or $h_y^{\rm rf}$ components can lead effectively to $\Delta m = 0$ transitions, which can mask $V_{\rm PNC}$ and cause an unwanted background.  
The net magnetic dipole transition amplitude, including the effect of $h_z^{\rm rf}$ and the transverse static fields, can be shown to be
\begin{equation}\label{eq:V_m2}
   V_m = -\frac{\sqrt{16-m^2}}{8} g_s \mu_B \mu_0 \:  \left\{ h_z^{\rm rf} + \left( \frac{h_x^{\rm rf} B_x + h_y^{\rm rf} B_y}{B_z} \right) \right\}.  
\end{equation}
We will return to these M1 contributions to the signal in Section \ref{sec:rfcavity}.  

We also consider possible interference to the signal from $\Delta m = 1$ transitions. The full-width-at-half maximum (FWHM) of the ground state  transition is $\sim 30$ kHz and the frequency of the $6s_{1/2} \: F=3, \: m=3 \rightarrow 6s_{1/2} \: F=4, \: m=4$ transition is detuned by $\sim 2.5$ MHz, or about 100 linewidths, from the rf frequency when $B_z = 7$ G. In addition, the spectral intensity of the rf signal 2.5 MHz from line center is about -100 dBc$/ \sqrt{Hz}$, as specified for our signal generator and amplifiers. These figures indicate that the rate of this $\Delta m = 1$ transition is exceptionally low.  In addition, these transitions, when they do occur, are expected to only add to the dc atomic signal, not to the sinusoidal modulation.  (That is, this signal cannot depend on the phase difference $\Delta \phi$ between the rf field and the Raman beams as the primary signal does.)  This argument is relaxed somewhat in the event that the Raman beams also excite any $\Delta m = 1$ transitions, but these transitions are also extremely rare, as the spectral density in the wings of the beat signal is -50 dBc$/ \sqrt{Hz}$. For these reasons, we expect the interference from $\Delta m = 1$ transitions to be negligible.

Next we consider the magnitude of Stark-induced transitions (that is, transitions allowed due to the presence of a static electric field $\mathbf{E}_0$ in the interaction region), and show that these are inconsequential relative to $V_{\rm PNC}$. On a $\Delta F = \pm 1$ transition, such as the transition considered in this work, the strength of the Stark transition is quantified by the vector polarizability $\beta$.  A sum over states expression for $\beta$ can be found in Refs.~\cite{BouchiatB75,GilbertWW84}.  Retaining only the $6p_{1/2}$ and $6p_{3/2}$ intermediate states, this polarizability reduces to 
\begin{equation}
    \beta \sim \frac{-1}{6} \left[\frac{| \langle 6p_{1/2}||r||6s_{1/2}\rangle |^2}{\left( E_{6s_{1/2}} -  E_{6p_{1/2}} \right)^2} + \frac{1}{2} \frac{| \langle 6p_{3/2}||r||6s_{1/2}\rangle |^2}{\left(E_{6s_{1/2}} -  E_{6p_{3/2}} \right)^2} \right] \times \Delta E_{6s_{1/2}, \rm{ hfs}},
\end{equation}
where $\langle 6p_j||r||6s_{1/2}\rangle$ are the reduced E1 matrix elements for the $6s_{1/2} \rightarrow 6p_j$ transitions, $E_{6s_{1/2}}$ and  $E_{6p_{j}}$ are the state energies, and $\Delta E_{6s_{1/2}, \rm{hfs}}$ is the hyperfine splitting of the $6s_{1/2}$ state. Using this expression, we estimate that the Stark polarizability for the hyperfine transition is $\beta = -0.0035 \; a_0^3$, a factor of $\sim 10^4$ smaller than $\beta$ for the $6s_{1/2} \rightarrow 7s_{1/2}$ transition.  Using a very pessimistic value of 0.1 V/cm for the static electric field $E_0$ inside the rf cavity, the ratio of the Stark amplitude to the PNC amplitude is at most
\begin{equation}\label{eq:Starkratio}
   \frac{\beta E_0}{\mathcal{E}_{PNC} }  \sim 0.033
\end{equation}
Thus the Stark-induced transition is not expected to be observable in the measurement of $\mathcal{E}_{PNC}$.

We close this section with an estimate of the magnitude of an electric quadrupole (E2) transition between ground state hyperfine components.  When the effect of hyperfine mixing is ignored, the E2 transition is forbidden since the triangle inequality ($J + J^{\prime}$ must be greater than or equal to two) is not satisfied. However, hyperfine mixing can weakly allow this interaction~\cite{Bouchiat1988}.  Derevianko~\cite{Derevianko16} has calculated the permanent electric quadrupole moment for each ground state hyperfine component and reported $Q_{F=3}^{\rm HFI} = 8.4 \times 10^{-6} \: ea_0^2$ and $Q_{F=4}^{\rm HFI} = -1.6 \times 10^{-5} \: ea_0^2$.  Estimating the transition quadrupole moment to be of this order, $Q \sim 1 \times 10^{-5} \: ea_0^2$, leads to 
\begin{equation}
    \frac{A_{E2}}{A_{PNC}} \sim \frac{Q \: (\partial e_z^{\rm rf}/\partial z)} {\mathcal{E}_{\rm PNC} e^{\rm rf}} \sim \frac{Q}{\mathcal{E}_{\rm PNC} \Delta z} 
 \sim 0.3
\end{equation}
where $\Delta z \sim 1$ mm is the distance over which the cavity field $e_z^{\rm rf}(z)$ turns on as the atoms enter or leave the cavity, as illustrated in Fig.~\ref{fig:Electricfieldrod}(c). Furthermore, the E2 interaction is active only over this short interval $\Delta z$, whereas the PNC interaction is applied over the entire $d =1.25$ cm width of the cavity.  Finally, the E2 contribution changes sign when reversing $m =3 \rightarrow m = -3$, whereas the E1 PNC contribution does not~\cite{zare1988book}.  Therefore, any residual E2 contributions can be eliminated by measuring the PNC signal on both $m$-states and using the average value.

Based on these analyses, we conclude that the primary challenge to a measurement of $\mathcal{E}_{\rm PNC}$ is the potentially large magnetic dipole contribution to the transition, and that taming these contributions will require extremely fine and precise tuning of the static magnetic flux density throughout the interaction region.

\section{The cylindrical rf cavity and its TM$_{010}$ mode}\label{sec:rfcavity}

\begin{figure}[b!]
 \begin{centering}
 	  \includegraphics[width=0.45\textwidth]{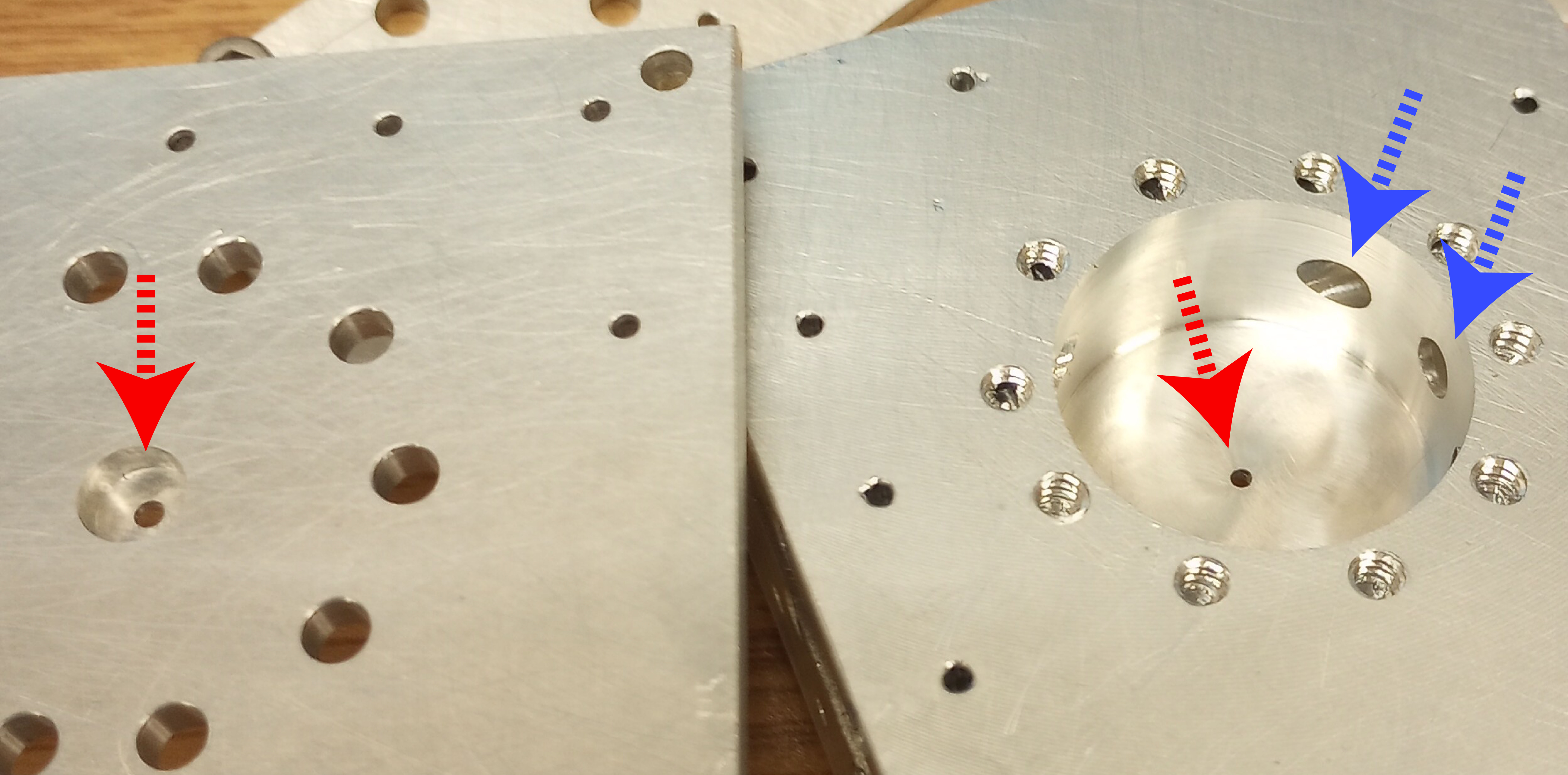}\\
  \caption{ Photograph of the machined base of the rf cavity assembly (right) and lid (left) that, when bolted over the base, forms the finished cavity. Atoms entering the science chamber through the recessed 1.5 mm diameter hole in the lid travel along the cavity axis, and exit through a similar hole in the base (red arrows). The adjacent excitation chambers cannot be seen in this view, but two of the coupling channels to the excitation chamber are visible (blue arrows). }
        \label{fig:machinedrfcavity}
 \end{centering}
\end{figure}

The entire rf cavity consists of a main cylindrical chamber located between two adjacent excitation chambers, designed to transfer rf power to the cylindrical chamber. The entire cavity was machined from a single block of aluminum with a separate removable lid. A photograph of the open cavity and lid is shown in Fig.~\ref{fig:machinedrfcavity}.
The excitation chambers are not visible in this image. The main cylindrical chamber (shown) was machined to a tolerance of 0.001" while the rest of the cavity was allowed a greater tolerance of 0.005".
The main cylindrical chamber is designed to resonate on the TM$_{010}$ mode at about 9.2 GHz. We will henceforth call this chamber the science chamber. 
The atoms enter the science chamber via the central atom hole (1.5 mm diameter) seen in the lid on the left of Fig.~\ref{fig:machinedrfcavity}, travel along the axis of the cylinder, and exit through a second on-axis atom hole in the body of the science chamber shown on the right in Fig.~\ref{fig:machinedrfcavity}. 
On the outside, we recessed both atom holes to help ease the machining of the rf cavity assembly and to reduce the length of the channel through which the atoms pass as well as any unwanted effects from it.
The resonant frequency $f_c$ of the rf cavity assembly matched the frequency $f_0$ of the $\Delta m=0$ $\psi_g \rightarrow \psi_e$ 
transition with an applied dc magnetic field of $\sim$ 7 G. (We did not attempt to tune the resonant frequency of the machined rf cavity, to avoid perturbing the cavity mode.)
The depth of the science chamber is $d=1.25$ cm, chosen to separate the frequencies of other cavity modes from that of the TM$_{010}$ mode, while also allowing adequate power levels of the TM$_{010}$ mode and a sufficient interaction length for the atom beam. 
We used numerical simulations to finalize the dimensions of the cavity and optimize the excitation of the mode.

In order to excite the TM$_{010}$ mode and to keep the mode as pure as possible, we used two adjacent excitation chambers to feed rf power into the science chamber, rather than exciting it directly. Aided by numerical simulations, we adjusted the dimensions of the excitation chambers to optimize the coupling of the rf power into the science chamber, with the excitation chamber at 1.65 cm $\times$ 0.5 cm $\times$ 4 cm and three coupling channels.
The two excitation chambers are positioned on opposite sides of the science chamber, each with three 4 mm diameter coupling channels leading to the science chamber, that are spaced by 8 mm. (Two of these coupling channels are visible in Fig.~\ref{fig:machinedrfcavity}.)  The excitation chambers are each excited via low-loss ($\sim$ 1 dB/m at a frequency of $f=9.2$ GHz) co-axial transmission lines, with the central pin of an SMA connector protruding into the chamber.
This cavity design offers several advantages, such as a uniform electric field amplitude across the interaction region, confinement of the electric and magnetic fields to a well-defined spatial region, and stronger field enhancement, when compared to an open parallel-plate transmission line structure that we explored previously~\cite{choipnc2016}.

 \begin{figure*}
 \begin{centering}
 \includegraphics[width=0.88\textwidth]{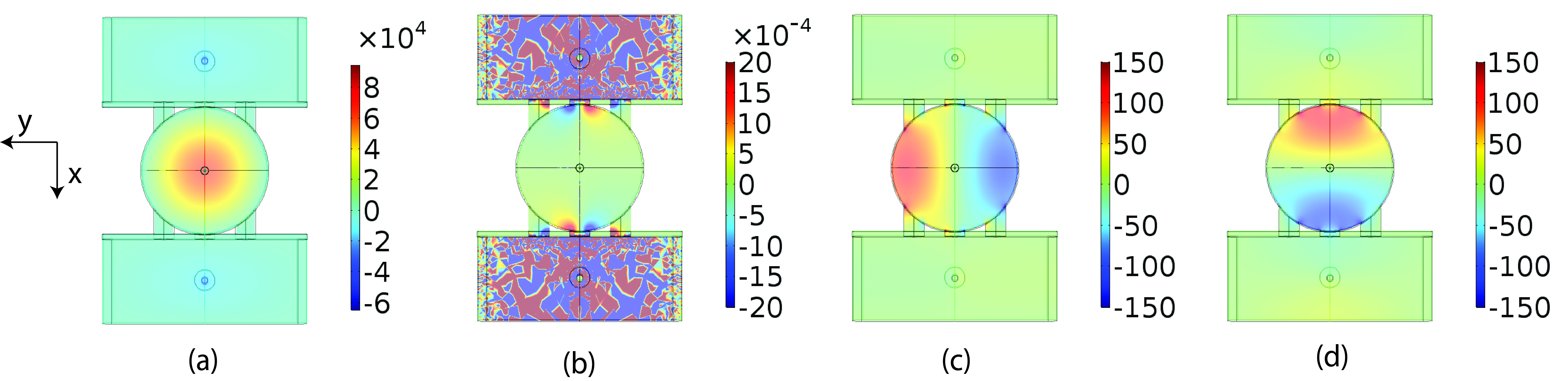} 
  \caption{Numerical simulation results of (a) $e_z^{\rm rf}$ (in V/m), (b) $h_z^{\rm rf}$ (in A/m), (c) $h_x^{\rm rf}$ (in A/m), and (d) $h_y^{\rm rf}$ (in A/m) in the rf cavity assembly.  The circular region is the science chamber, while the two rectangular regions (above and below the science chamber in these diagrams) are the excitation chambers. Color bars indicate the scale for each field amplitude. The atom beam propagates into the page (z-direction) at the center circle in the science chamber.}
        \label{fig:comsolall}
 \end{centering}
 \end{figure*}
The cavity mode is illustrated in Fig.~\ref{fig:comsolall}, which shows color maps of (a) $e_z^{\rm rf}$, (b) $h_z^{\rm rf}$, (c) $h_x^{\rm rf}$, and (d) $h_y^{\rm rf}$ in the rf science chamber and excitation chambers. As discussed earlier, $e_z^{\rm rf}$ is strongest at the center of the cavity ($\rho = 0$), and decreases to zero at the walls.  With an input power of 1 W on each of the two excitation ports, the on-axis electric field amplitude is $e_0^{\rm rf} = 72$ kV/m. At this field strength, and using the expected weak-force-induced moment $\mathcal{E}_{\rm PNC} = i \kappa \: 1.82 \times 10^{-11} \: e a_0$~\cite{DzubaBFR12} with $\kappa = 0.117$, the Rabi frequency of the interaction with the rf field is $\Omega_{\rm PNC} = 0.0123$ rad/s. An average velocity (270 m/s) \cite{ramsey1956molecular} atom transits the  science chamber in time $\tau = d/v = 44 \ \mu$s, resulting in a Bloch precession angle of $\Omega_{\rm PNC} \ \tau \sim 0.54 \; \mu$rad.  

The cavity mode also includes transverse magnetic field components $h_x^{\rm rf}$ and $h_y^{\rm rf}$, as discussed in Sec.~\ref{sec:interactions} and shown in Fig.~\ref{fig:comsolall}(c) and (d),
which contribute to the $\Delta m = 0$ signal if $B_x$ and $B_y$ are not sufficiently zeroed.
To this point, we have reduced $B_x$ and $B_y$ to less than 30~mG throughout the interaction volume, while $B_z \sim 7$~G.  
With 1~W of power input to each of the two cavity inputs, numerical simulations show that the magnetic field amplitude of $h_{\phi}^{\rm rf} $ at a radius of 0.5 mm (that is, at the edge of the atomic beam), is $\sim$10 A/m.  (The magnetic field $h_{\phi}^{\rm rf} $ is zero at the center of the cavity.) 
Using these parameters, the ratio of the maximum magnetic dipole signal to electric PNC dipole signal is $\sim2 \times 10^4$.
As will be discussed in Sec.~\ref{sec:popmod}, this magnetic dipole signal is greatly reduced by spatial averaging over the cross section of the atomic beam. 

\begin{figure}[b!]
 \begin{centering}
  \includegraphics[width=0.45\textwidth]{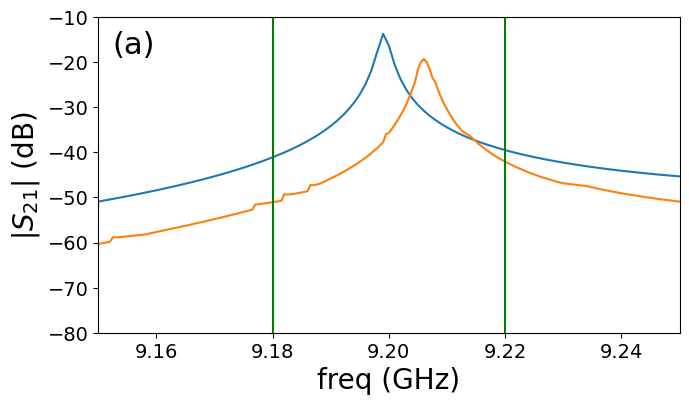} \\
  \includegraphics[width=0.45\textwidth]{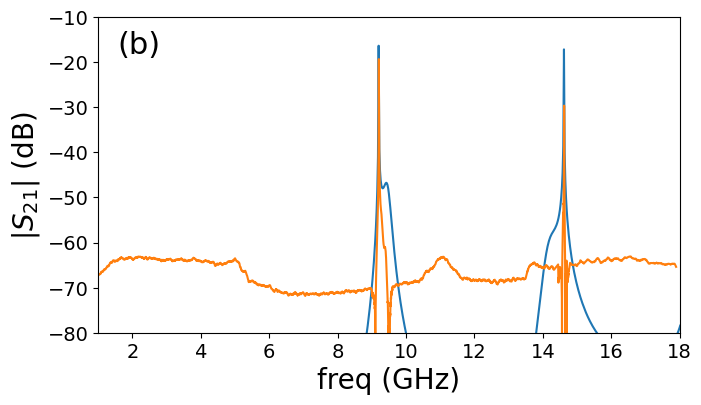}
 
   \caption{ The $S$-parameter $\lvert S_{21}\rvert$ vs.\ frequency for the rf cavity assembly. The orange trace shows the measurement, while the result of numerical simulations is shown in blue.  (a) The narrow spectrum featuring the TM$_{010}$ peak centered at 9.2 GHz. The two green lines show the range of frequencies corresponding to variations in the cavity radius $a$ differing by $\pm$ 0.001''.  (b) A broader spectrum illustrating the clear separation between the TM$_{010}$ peak at 9.2 GHz and the closest mode, the TM$_{110}$ peak at 14.7 GHz.}
         \label{fig:finefreqcompare}
\end{centering}
\end{figure}

The measured resonant frequency of the rf cavity is $f_c =$~9.207~GHz. This is in good agreement with the simulated resonant value of 9.200 GHz, but differs slightly from the resonant frequency $c p_{0,1}/2 \pi a =$ 9.246 GHz of the ideal cylindrical cavity. 
The difference between these frequencies is due to the atom holes and power coupling channels in the cavity, and the machining tolerance of the cavity radius $a$. 
 The FWHM of the cavity resonance is approximately 2 MHz and its Q is $\sim$ 4500. See the S-parameter plot ($\lvert S_{21}\rvert$, the orange trace) in Fig.~\ref{fig:finefreqcompare}(a). The peak frequency, bandwidth, and bandshape of the measured $\lvert S_{21}\rvert$ parameter are closely matched by those of the numerical simulation (blue curve). The difference in peak heights could be due to losses in transferring power into the science chamber, which seem to be very sensitive to the precise diameter of the power coupling channels.
The plot in Fig.~\ref{fig:finefreqcompare}(b) shows $\lvert S_{21}\rvert$ over an expanded frequency range.  Again, the orange trace is the measured $S$-parameter, while the blue trace is the simulation result. Besides the peak corresponding to the TM$_{010}$ (9.2 GHz), a feature due to the TM$_{110}$ (14.73 GHz) is also visible.  

\section{Experimental geometry}\label{sec:Exp} 

\begin{figure}
    \centering
    \includegraphics[width=0.4\textwidth]{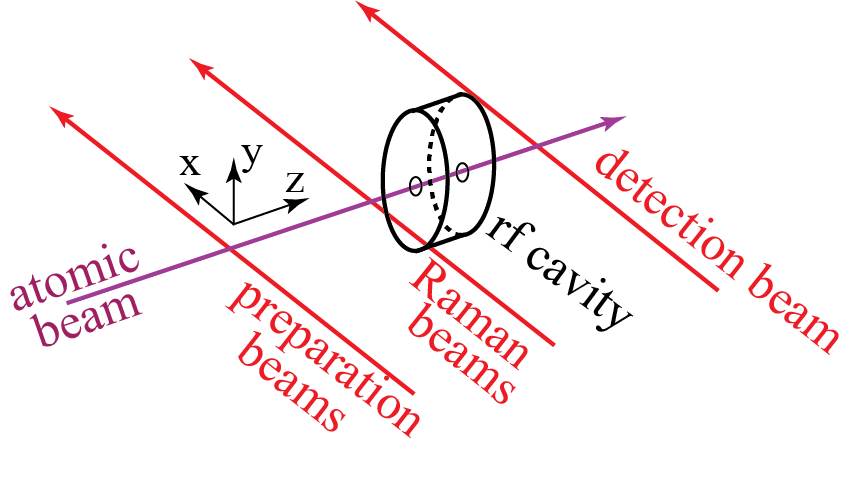}
\caption{Experimental configuration. A beam of cesium atoms is generated in an effusive oven.  The atoms are optically pumped into the $\psi_g$, the $F=3$, $m$ (where $m=\pm3$) hyperfine component of the ground $6s_{1/2}$ state by the preparation beams, and excited into a superposition state $(\psi_g + \psi_e)/\sqrt{2}$ by the Raman beams. They interact with the TM$_{010}$ mode in the rf cavity, and the population of $\psi_e$, the $F=4$, $m$ hyperfine component of the $6s_{1/2}$ state measured using the detection beam.}
  \label{fig:exp_layout}
\end{figure}

The measurement is set up in an atomic beam system, most of which has been described previously~\cite{antypasm12013}. A simplified diagram of the experimental configuration is shown in Fig.~\ref{fig:exp_layout}. 
Atoms effusing from a heated cesium oven through a nozzle ($\phi =1$ mm diameter) form a horizontal atomic beam. These atoms are optically pumped into the $(F, m) = (3,3)$ or $(3, -3)$ hyperfine component of the ground state by a pair of $\lambda = 852$ nm laser beams.  Raman spectra indicate that approximately 80\% of the atoms populate this initial state.  
A 1 mm diameter aperture positioned 30 cm from the oven nozzle, but before the rf cavity assembly, further defines the atom beam in the science chamber. 
Because of the wide distribution of longitudinal velocities in the atomic beam, the range of interaction times $\tau = d/v$  of the atoms with the rf field inside the science chamber is also broad~\cite{ramsey1956molecular}. Our measurements show that low velocity atoms, which are expected to preferentially fall out of the atom beam due to gravity, do not contribute to the atomic signal.

Population entirely in one state or the other (that is, $\psi_g$ or $\psi_e$) is represented by a Bloch vector pointing along the $\pm z$-axis, and an interaction with a near resonant field causes a rotation of the Bloch vector on the surface of the Bloch sphere~\cite{foot2005atomic}. Since the change in the projection of the Bloch vector on the $z$-axis is maximized when the Bloch vector is initially in the $x-y$ plane, we prepare the atoms in a superposition state $(\psi_g + \psi_e)/\sqrt{2}$ before they enter the  science chamber, using a pair of overlapping laser beams which drive a Raman transition between the two ground state levels.  These linearly-polarized ($\hat{z}$-direction), parallel laser fields are two-photon resonant with the $\psi_g \rightarrow \psi_e$ transition; each is detuned from resonance with the $6s_{1/2} \rightarrow 6p_{3/2}$ transition by $\Delta/2 \pi\sim 230$ MHz.  At this detuning, the Raman interaction is sufficiently strong to produce the desired superposition state, yet the ac Stark shift of the ground state transition is sufficiently small due to the opposing directions of frequency shift from the two different hyperfine ground state levels.  
The powers of the Raman beams are 0.4 mW and 2.3 mW, and the beam radius is 7.5 mm.  We maintain this large beam size to assure that the variation of the Raman beam intensities over the atomic beam cross section is small.  
We stabilize the power of the Raman beams using an acousto-optic modulator (AOM) and feedback circuit. We also stabilize the frequency of these lasers using a saturated absorption resonance in a tabletop Cs cell.  

The evolution of the superposition state of the atoms within the science chamber depends upon the phase difference, $\Delta \phi$, between the Raman interaction which prepares the initial superposition state and the rf interaction in the science chamber.  If these two interactions are in phase with one another, then the Bloch vector describing the superposition state continues to precess towards $\psi_e$, increasing the amplitude $c_e(t)$ and decreasing $c_g(t)$. (These amplitudes were defined in Eq.~(\ref{eq:superposition}).) Conversely, if the rf interaction is out of phase with the Raman interaction, then the population amplitude $c_e(t)$ decreases during the interaction with rf field. When the phase between the Raman and rf interactions is scanned linearly, the population of $\psi_e$ modulates sinusoidally, and the amplitude of this modulation provides a means of determining the magnitude $|V_{\rm PNC}|$ of the interaction term. 

Other field reversals can also be performed that can affect the interference. 
This is seen most directly by examining the field products that influence the evolution of the atomic Bloch vector. In our geometry, the Raman lasers excite a $\Delta m = 0$ transition (they are each linearly polarized in the $z$-direction), and the rf electric field is also polarized in the $z$-direction and excites a $\Delta m = 0$ transition.  The static magnetic field $\mathbf{B}$ defines the $z$ axis, leading to a parity non-conserving, time-reversal symmetric product of interaction terms of the form $i (\mathbf{e}^{R1} \cdot \mathbf{e}^{R2}) (\mathbf{e}^{rf} \cdot  \mathbf{B})$. $\mathbf{e}^{R1}$ and $\mathbf{e}^{R2}$ are the electric field amplitudes of the two Raman beams.  The phase of this term can be varied by invoking a phase shift between the two Raman beams, by reversing $\mathbf{B}$, or by varying the phase difference $\Delta \phi$. The first two operations can be used as a system check, but they are not an integral part of the PNC measurement.  Rather, variation of the phase difference $\Delta \phi$ between Raman and rf interaction is all that is necessary.  We discuss our technique for this phase control in section~\ref{sec:popmod}.

The transition strength of the Raman interaction was one of our primary reasons for selecting a $\Delta m = 0$ transition for the measurement of $\mathcal{E}_{PNC}$.  
If the polarization of one of these Raman beams were rotated $90^{\circ}$, the Raman interaction would drive a $\Delta m = \pm 1$ transition.  The amplitude of this transition is very weak, however, due to cancellation between contributions to the two-photon moment through $F=3$ and $F=4$ intermediate states (hyperfine components of the $6p_{3/2}$ state), which are of similar amplitude but opposite in sign.  This cancellation becomes less complete with smaller detunings from the intermediate resonance, but at the expense of large ac Stark shifts of the levels.  For this reason, we chose to work with a $\Delta m = 0$ transition for the measurements.  

After the atoms interact with the rf field in the cavity, we detect the population of the $F=4$ ground state by driving a cycling transition in the detection region from this level to the $6p_{3/2} \: F=5$ level, and collect fluorescence on a large area photodetector. This photocurrent is then amplified in a transimpedance amplifier with a gain of 20 M$\Omega$. The signal is further amplified by a factor of ten in a simple op-amp circuit, and then input to a commercial lock-in amplifier.

Careful control of the static magnetic field $\mathbf{B}$ in the optical pumping, interaction, and detection regions is essential to isolating $V_{\rm PNC}$ from $V_m$. We implement this control using magnetic field coils inside and outside the vacuum system.  Three pairs of large coils around the aluminium vacuum chamber minimize the Earth's magnetic field in the interaction region (two coils for each axis). Inside the vacuum chamber there are 17 additional rectangular coils, separated into three main regions, whose purpose is to either install a particular direction of magnetic field in the region or correct a magnetic field gradient in the region.

To set the correct currents in the different magnetic field coils in the various regions, we observed various features of the spectra, either with only the rf source active or with the rf and Raman lasers active, and optimized those features.  With $B_z$ adjusted slightly to bring the $\Delta m = +1$ transition into resonance with the rf field, the rf power reduced, and the Raman beams blocked, we zeroed the gradient of $B_z$ by minimizing the width of this peak, typically from $\sim$10 to 40~kHz FWHM, depending on rf power. 

We also use this $\Delta m = +1$ resonance to adjust the relative phase of the rf inputs to the two excitation chambers.  When the two inputs are in phase with one another, the cavity mode field amplitude is maximized, as is the magnitude of the rf spectrum. We use equal length transmission lines to carry the signal to each input, and fine tune the phase using trombone line phase shifters.

\begin{figure}
 \begin{centering}
 	  \includegraphics[width=0.45\textwidth]{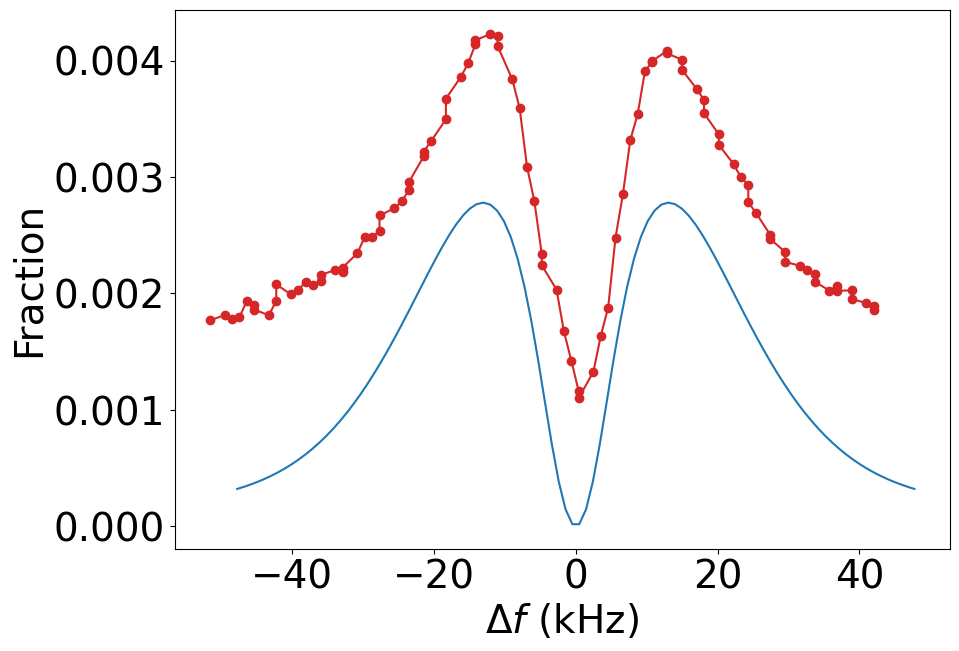}\\

  \caption{ Spectra of the $\Delta m = 0$ (red curve) peak with only the rf field applied and $B_x$ and $B_y \ll B_z$. The blue curve is the result of numerically integrating Eqs.~(\ref{eq:equationofmotionexcited}) and (\ref{eq:equationofmotionground}), setting $B_x=0$, and letting $B_y$ vary linearly from -30 mG to +30 mG across the science chamber. This curve represents the spatial average over the cross section of the atom beam. 
    } 
  \label{fig:zerob}
 \end{centering}
 \end{figure}

We then adjusted $B_z$ to bring the $\Delta m = 0$ peak back into resonance, and collected spectra such as that shown by the trace in Fig.~\ref{fig:zerob}. The abscissa in this plot is the detuning $\Delta f = f-f_0$, where $f_0$ is close to the cavity resonance frequency $f_c$. To collect these spectra, we tuned through the resonance by tuning $B_z$, thus shifting the resonant frequency of the atoms. (We chose this method instead of tuning the rf frequency to avoid changes to the cavity mode pattern or amplitude excited by the rf input at different frequencies.) 
Spectra such as the one shown in Fig.~\ref{fig:zerob} help us to zero the static transverse fields $B_x$ and $B_y$, since, with only the rf field applied, the $\Delta m =0$ magnetic dipole signal should reduce to zero when $B_x$ and $B_y$ are zero. 
Under these conditions, the only signal remaining is due to $V_{PNC}$, which should be too tiny to observe with only the rf field applied.
Due to non-uniformities in $B_x$ and $B_y$\ (currently estimated to vary from -30 mG to +30 mG across the science chamber), however, we could only null the signal at line center, and the signal persisted at frequencies above and below resonance, as shown in the trace in Fig.~\ref{fig:zerob}. In the future, better suppression of $B_x$ and $B_y$ throughout the  science chamber will be necessary.

To calibrate the measurement, we must determine the scaling between the fractional change in population and the signal measured by the detection laser and photodiode. To this end, we measure the atomic signal when the preparation lasers are turned off, and also when the preparation lasers are on.  In the former, the atomic beam is unpumped, and the detectable population is 9/16 of all atoms, reflecting the degeneracy of the $F=4$ and $F=3$ states, plus background signal from scattered light.  In the latter, the $F=4$ level is nearly empty, and only the background signal remains.  The difference between these two measurements, multiplied by a factor 16/9, represents the signal when the entire population has been promoted to the $F=4$ level. The fraction of atoms excited, as shown as the ordinate in Fig.~\ref{fig:zerob}, is the ratio of atoms in the excited state relative to the total number of atoms in the system.

We {also require a means of calibrating the rf field amplitudes inside the cavity. To achieve this, we tuned the $6s_{1/2} \ F=3, m=3 \rightarrow 6s_{1/2} \ F=4, m=4$ transition back into resonance with the rf field (by decreasing the Zeeman field). As a $\Delta m = +1$ transition, this transition is driven by the $h_x^{\rm rf}$ and $h_y^{\rm rf}$ field components of the TM$_{010}$ mode.  We show the plot of the population transferred to the $6s_{1/2} \ F=4, m=4$ state as a function of rf power in Fig.~\ref{fig:rfsweep}.
\begin{figure}[b!]
 \begin{centering}
 	  \includegraphics[width=0.45\textwidth]{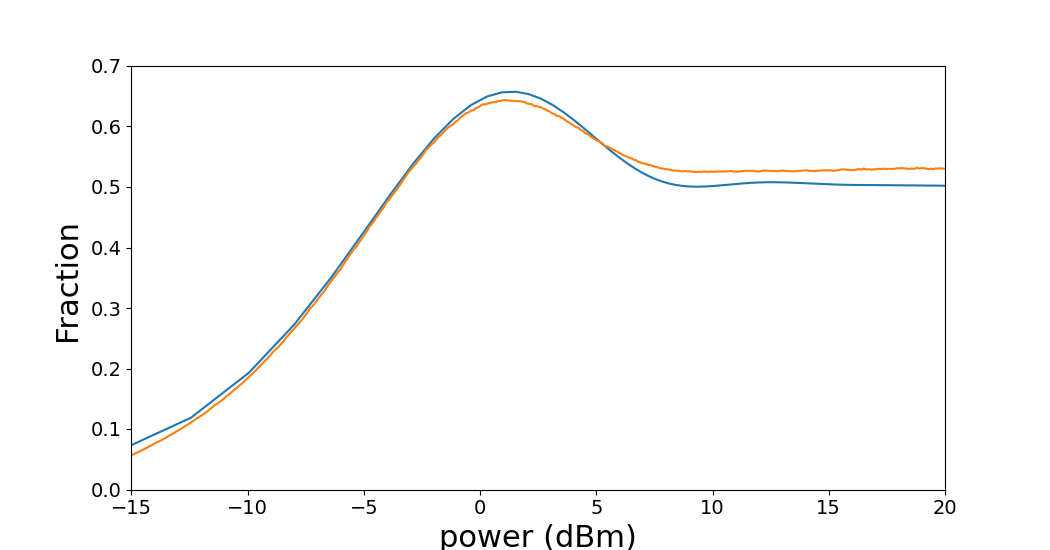}\\

  \caption{The fraction of atoms transferred from $\psi_g = \  6s_{1/2} \ F=3, m=3$ to $6s_{1/2} \ F=4, m=4$  by the rf field, as a function of the rf power. This plot shows the experimental data (orange) and the numerical integration results (blue).
}
        \label{fig:rfsweep}
 \end{centering}
 \end{figure}
This plot shows the experimental data (orange) and the numerical integration results (blue).
A population fraction of one signifies that all of the population that was initially in the $6s_{1/2}\ F=3, m=3$ state is transferred to the $6s_{1/2}\ F=4, m=4$ state.  
For the calculated curve, we used the $h_x^{\rm rf}$ and $h_y^{\rm rf}$ determined through the numerical calculation of the cavity mode. The calculated curve agrees well with the measurements after a 3 dB adjustment of the rf power, which indicates that the power coupled into the science chamber is somewhat smaller than simulated. This result is consistent with Fig.~\ref{fig:finefreqcompare}(a), which shows that the parameter $|S_{21}|$ is $\sim$5 dB smaller than calculated at 9.2 GHz.

\section{Population modulation}\label{sec:popmod}

To carry out these measurements, phase coherence between the rf cavity mode and the Raman interaction that prepares the atoms is essential. This coherence is accomplished using a pair of signal generators (143 MHz for one, 9.2 GHz for the second), that are phase locked to one another via their internal 10 MHz reference clocks, as shown in Fig.~\ref{fig:rfgeneration_technique}. 
\begin{figure}
 \begin{centering}	\includegraphics[width=0.45\textwidth]{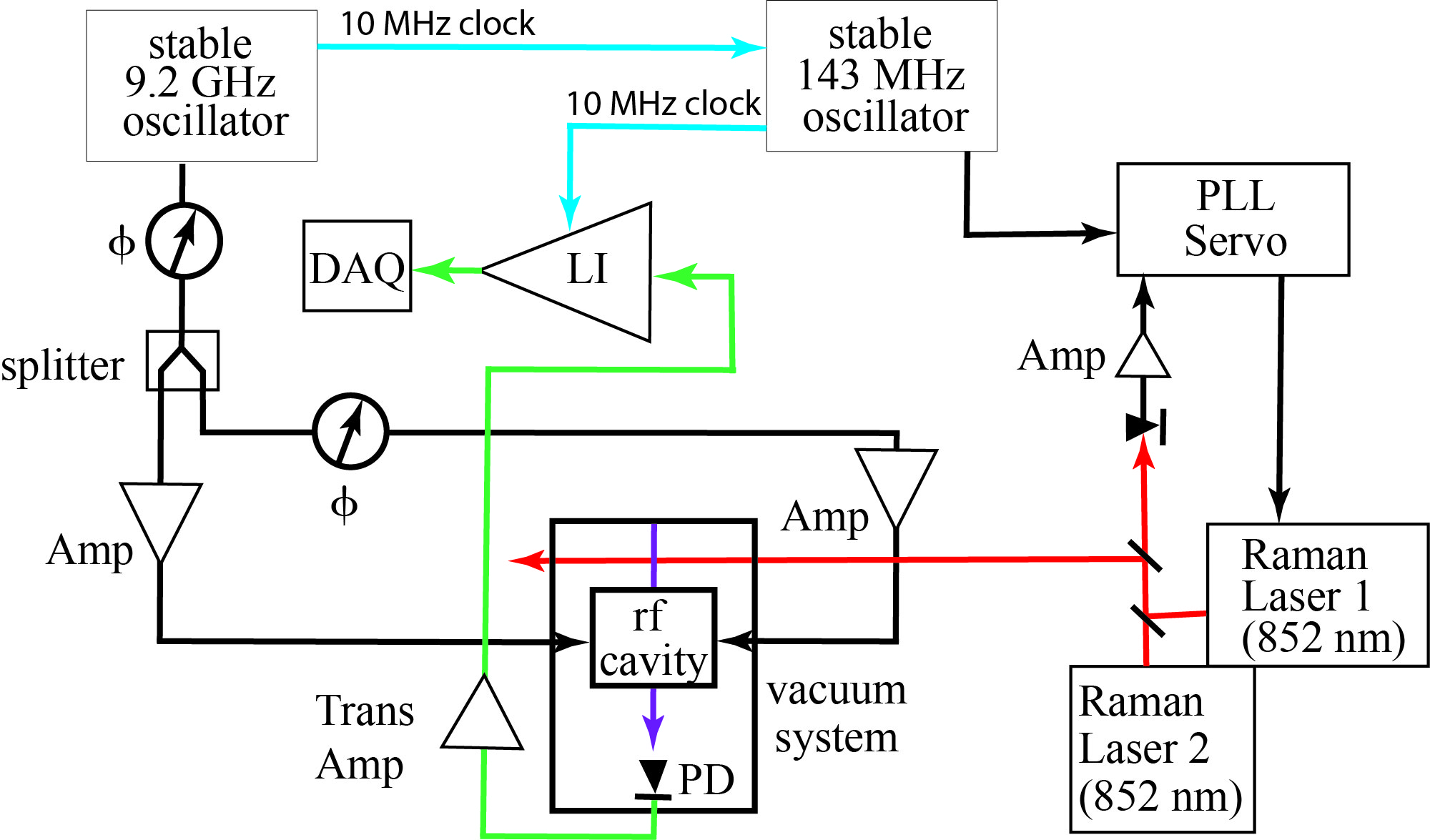}
\caption{Simplified diagram depicting the phase coherent rf generation technique.  The Raman laser beams (red traces) prepare the ground state atoms in the superposition state $(\psi_g + \psi_e)/\sqrt{2}$.  The rf field drives a ground state transition $\psi_g \rightarrow \psi_e$, and we detect the small change in population of state $\psi_e$.
Coherence between the Raman and rf interactions is maintained using two highly stable rf oscillators with a common 10.0 MHz reference clock (blue traces).  PD = photodiode, $\phi$ = phase shifter, PLL = phase-lock loop , LI = lock-in amplifier, DAQ = data acquisition system.}
    \label{fig:rfgeneration_technique}
 \end{centering}
\end{figure}
We beat the Raman laser beams on a fast Schottky photoreceiver and amplify the beat note to be accepted by a commercial phase-lock loop (PLL) servo (Vescent D2-135). Internally, the PLL divides down the beat note frequency by a factor of 64, and compares the divided beat note's phase with the reference signal, which originates from the 143 MHz signal generator. This servo phase locks the frequency difference between the two Raman lasers to the signal generator, maintaining resonance with the ground state $\psi_g \rightarrow \psi_e$ transition at 9.2 GHz. The second signal generator is then used to generate the 9.2 GHz signal directly, which is split into two lines, amplified, and applied to the two input ports on the rf cavity assembly. Since the references of each generator are locked together, the offset phase of the lasers is coherent with the 9.2 GHz source driving the rf cavity assembly inputs. 

We show in Fig.~\ref{fig:ramanrfintereferncespecturm} the Raman spectrum (orange trace), the rf spectrum (blue trace), and the spectrum when both the rf and Raman interactions are turned on (green trace), illustrating interference fringes near the peak resulting from the coherence of the two interactions. The phase difference between the two interactions is held constant while the frequency is ramped to obtain this spectrum.
The amplitude of these fringes is much larger than the rf signal alone, as expected.  
\begin{figure}
 \begin{centering}
 	  \includegraphics[width=0.45\textwidth]{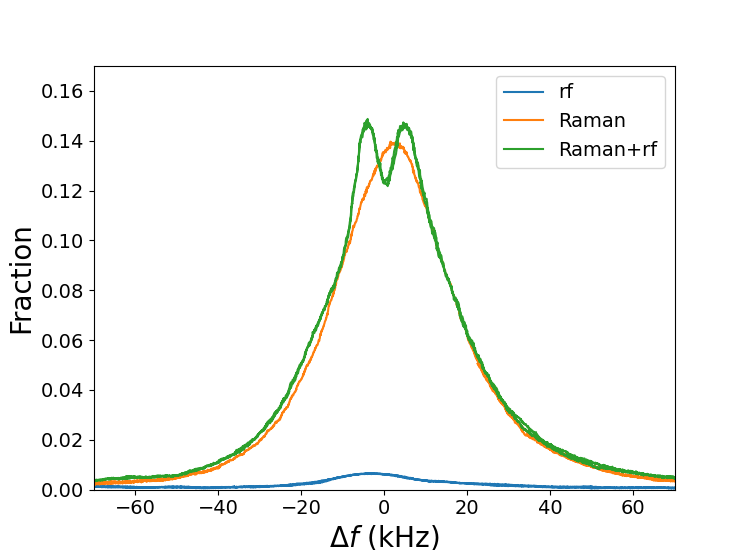}\\

   \caption{Spectra showing the fraction of the population excited to the $\psi_e = \  6s_{1/2}, (F,m)=(4,3)$ state vs.\ the frequency detuning from resonance $\Delta f = f-f_0$. The traces are the Raman spectrum (orange), the rf spectrum (blue), and the spectrum when both the rf and Raman interactions are turned on (green). A relatively large transverse magnetic field ($B_x$ and $B_y$) is applied so that the magnetic interaction $V_m$ is strong. 
   (The frequency shift between the Raman and rf resonances is the result of a gradient in $B_z$ that accompanied the large transverse magnetic field.)   }   \label{fig:ramanrfintereferncespecturm}
  \end{centering}
\end{figure}
(The peaks of the Raman and rf spectra are offset slightly from one another due to a gradient in $B_z$ that accompanied the large transverse magnetic field.)

In Fig.~\ref{fig:ramanrfinterefernce}, we show the sinusoidal variation of the signal while scanning the phase difference $\Delta \phi$ between the rf and Raman interactions near the peak of the spectrum for several values of $V_m$. We control the magnitude and sign of $V_m$ by changing the transverse magnetic field $B_x$ and $B_y$. (See Eq.~(\ref{eq:V_m2}).)  In Fig.~\ref{fig:ramanrfinterefernce}, for small $V_m$ (blue trace), the signal modulation is small, while for large $V_m$ (red and green traces), the signal modulation is large.  The red and green traces show a 180$^{\circ}$ phase shift from one another, as $V_m$ is of opposite sign for these two traces. 

Since the rf magnetic field $\mathbf{h}^{\rm rf}$ in the cavity is azimuthal, the interaction $V_m$ for atoms on one side of the cavity axis is of opposite sign to $V_m$ on the other. (See the magnetic dipole interaction term in Eq.~(\ref{eq:V_m2}), particularly the terms involving $h_x^{\rm rf}$ and $h_y^{\rm rf}$.) The local signal modulations on the two sides of the beam are therefore $180^{\circ}$ out of phase with one another.  We expect that, with spatial averaging over the atom beam, this magnetic dipole signal will be strongly reduced.  
In order to take best advantage of this effect, we have mounted the rf cavity assembly on a multi-axis positioner, with spatial step sizes of less than 30 nm.  This positioner allows us to finely adjust the relative position of the atom beam within the rf cavity assembly, in order to minimize the transverse magnetic dipole contribution to the signal.
Cancellation through spatial averaging will be complete if the atom beam is perfectly centered on the science chamber axis.  If the atom beam is somewhat off center, we calculate that the average magnetic dipole signal, normalized to the PNC signal is roughly
\begin{equation}
    \frac{|V_m(\rho_0)| 4\rho_0 \Delta x}{|V_{\rm PNC}| \pi \rho_0^2 } \approx \frac{\sqrt{7}}{2\pi} \left(\frac{g_s \mu_B \mu_0}{\mathcal{E}_{\rm PNC}} \right) \left( \frac{h_y^{\rm rf}(\rho_0)}{e_z^{\rm rf}(0)} \right) \left( \frac{B_y}{B_z} \right)  \left( \frac{\Delta x}{\rho_0} \right),  
\end{equation}
where $\rho_0 = 0.5$ mm is the radius of the atomic beam, $h_y^{\rm rf}(\rho_0) \sim 10$ A/m is the magnetic field amplitude at a distance $\rho_0$ from the axis of the science chamber, $e_z^{\rm rf}(0) \sim 70$ kV/m is the electric field amplitude at or near the science chamber axis, and $\Delta x$ is the distance from the center of the atom beam to the science chamber axis. Using $B_y$ = 3 mG and $\Delta x = 30$ nm
(smallest resolution of the positioner),
we estimate the transverse magnetic signal is a factor $2$ times larger than the PNC signal. Note also that since $h_x^{\rm rf}$ and $h_y^{\rm rf}$ are $\pi/2$ out of phase with $e_z^{\rm rf}$, and $\mathcal{E}_{\rm PNC}$ is imaginary, $V_m$ and $V_{\rm PNC}$ are in phase with one another, these amplitudes add directly. This property increases the importance of reducing the magnetic dipole contribution through spatial averaging.  Additional reduction of the transverse magnetic fields $B_x$ and $B_y$ is required for a successful measurement of $\mathcal{E}_{\rm PNC}$.

\begin{figure}
 \begin{centering}
 	  \includegraphics[width=0.45\textwidth]{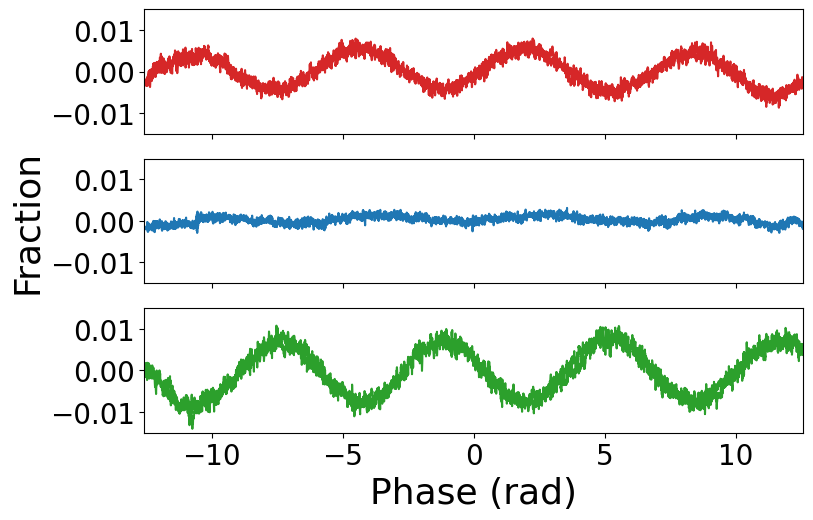}\\

   \caption{ Modulation of the fraction of the population in the $\psi_e = \  6s_{1/2}, (F,m)=(4,3)$ state vs.\ phase difference between the rf and Raman interactions. The rf interaction is primarily $V_m$, as defined in Eq.~(\ref{eq:V_m2}).  $V_m$ is quite small for the blue trace, for which $B_y = 0$ mG, but large for the red ($B_y=$ 125 mG) and green ($B_y=$ -125 mG) traces.  Note the $180^{\circ}$ phase shift between the red and green traces, consistent with sign change of $V_m$. $B_x=$ 75 mG for each trace.  This signal modulation (population fraction) sits on top of a large dc signal, of magnitude $\sim$0.5 population fraction. }   \label{fig:ramanrfinterefernce}
  \end{centering}
\end{figure}

We analyzed the noise level at different frequencies in our system to select the optimal modulation rate.
We found that spurious signals at 60 Hz and its harmonics were the worst offenders, along with low frequencies in general, as expected. Additionally the broad distribution of atomic velocities in the atom beam imposes an upper limit on the modulation rate.
With those facts in mind we chose 150 Hz as the phase frequency. At this frequency, the S/N ratio of our measurement is dominated by shot noise and Johnson noise. For a 1V signal size, there is $\sim$21 $\mu V$ of rms noise in a 1 Hz bandwidth at 150 Hz.

\section{Modulation Spectra\label{sec:ModSpec}}

 \begin{figure}
 \begin{centering}
 	  \includegraphics[width=0.45\textwidth]{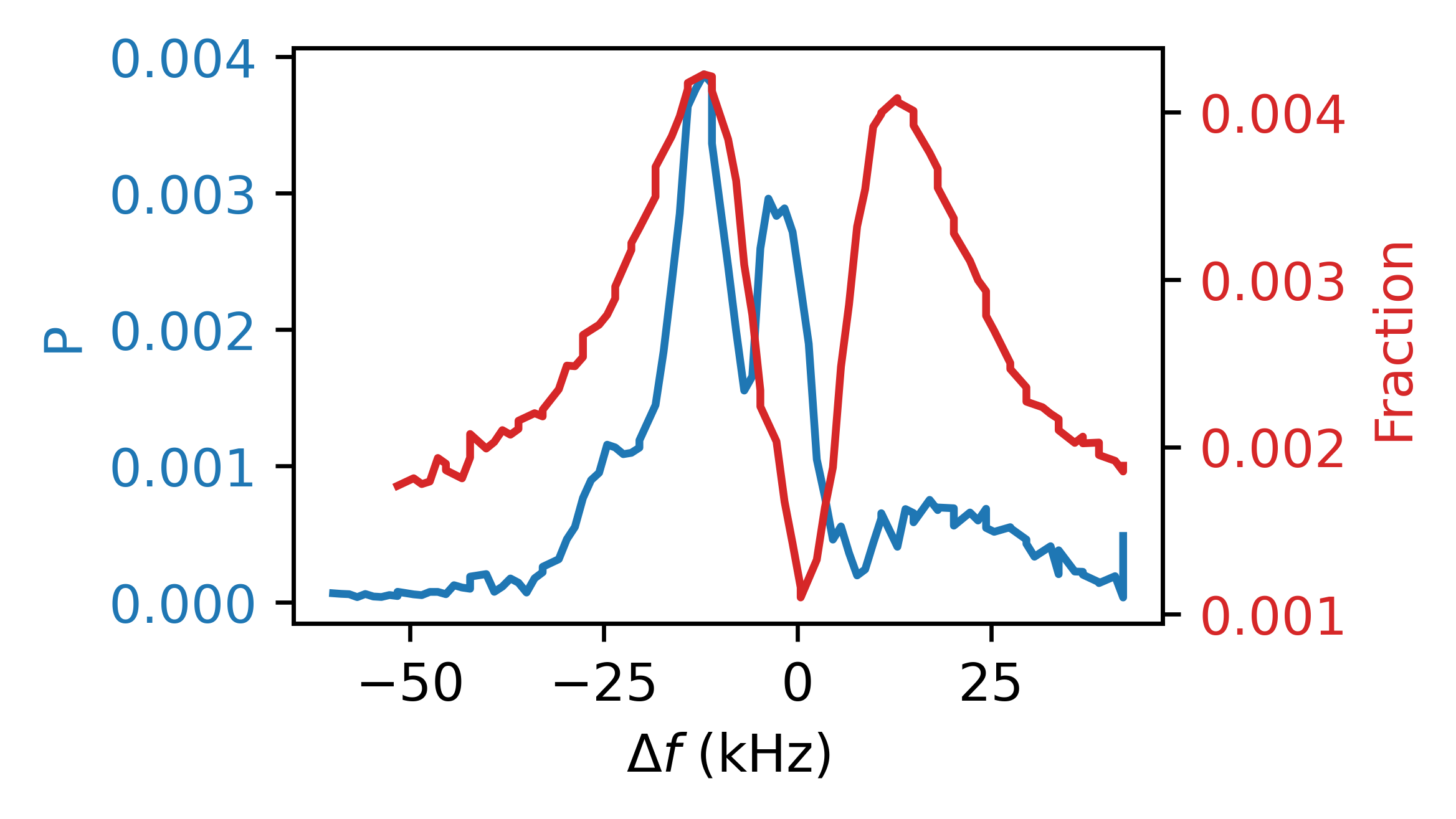}\\

  \caption{The spectrum $P$ of the population modulation (blue trace), as defined in Eq.~(\ref{eq:deltaP}), when both the Raman and rf fields are applied, as a function of the frequency detuning from resonance $\Delta f$.  The transverse magnetic field $B_x$ and $B_y$ are minimized in this plot.  For comparison, we reproduce the rf-only spectrum (red trace), in which the Raman interaction is removed, from Fig.~\ref{fig:zerob}.  }
  \label{fig:interrfa}
 \end{centering}
 \end{figure}
 
When we scan across the transition frequency with the rf and Raman fields applied, and scan the phase difference $\Delta \phi$ between the Raman and rf interactions, we found an interesting interference signal, as seen in the blue trace in Fig.~\ref{fig:interrfa}. In this spectrum, we plot $P$, the amplitude of the modulation of the $F=4$ population, 
\begin{equation}
     P=\frac{|c_e(max)|^2-|c_e(min)|^2}{2},
    \label{eq:deltaP}
\end{equation}
where $c_e(max)$ ($c_e(min)$) is the maximum (minimum) value of $c_e$ as the phase $\Delta \phi$ changes.  
These spectra were distinctly different from the spectra with only the rf applied, as expected.
We systematically varied different experimental parameters to understand the features of the spectra. Moving the atom beam relative to the cavity axis did not change the spectra. If the interference was mainly caused by $h_x^{\rm rf}$ and $h_y^{\rm rf}$ in the science chamber, we should see that moving the atom beam relative to the rf cavity assembly would change the signal as the spatial average of the signal across the mode pattern changed. Changing either the gradient or magnitude of $B_x$ or $B_y$ should have a similar effect, 
but we did not see a change in the general spectra. Changing $B_x$ or $B_y$ affected the middle of the rf-only scan (red); in the interference spectrum (blue), the peak at that same frequency increased as well. We also note the asymmetry in the interference spectrum, which we attribute to the gradients in the magnetic field.

We considered the role of 9.2 GHz rf magnetic fields outside the cavity, which could also potentially cause ground state transitions. Stray high-frequency fields surrounding co-axial connectors, and methods for their reduction, have been investigated previously~\cite{michalka2018}.  We expected {these contributions} to be small, but they required investigation. For this purpose, we computed the external fields numerically and found that external fields originate from the atom holes and from the co-axial cable connectors. We experimentally verified these numerical results and found that the strongest 9.2 GHz fields were near the cables
and cable connectors.  We were able to reduce its strength by up to two orders of magnitude by adding more shielding to the co-axial transmission lines that carry the rf power, and rf absorbers around the rf cavity assembly. After implementing these changes and installing the cavity inside the vacuum chamber, we saw no change to the interference signal.

Finally, we carried out a set of measurements in which we added a few cm of delay line to one of the transmission lines that drives the rf cavity assembly, introducing a $\pi$ phase difference between the inputs. We observed the excitation spectrum with only the rf inputs applied (i.e.~no Raman interaction), such as that shown in Fig.~\ref{fig:zerob}, while adjusting the phase delay, and minimized the excitation signal, consistent with the two inputs adding out-of-phase with one another. Under this same condition, but with the Raman lasers turned on, however, the magnitudes of the spectra with the two rf inputs in-phase or $\pi$ out-of-phase are quite similar, as seen in Fig.~\ref{fig:interrfb}. The frequencies of the peaks and valleys change, but their overall amplitudes are remarkably similar.
Our numerical simulations of the fields within the rf science chamber paint a consistent picture. 
With the rf inputs out-of-phase with one another, the amplitude of the  TM$_{010}$ mode is extremely weak.  These same simulations, however, indicate a non-negligible magnetic field component $h_z^{\rm rf}$ near the atom entry holes, which persists even when the phase of one of the rf inputs is shifted by $\pi$.

Based on these observations, we currently believe that $h_z^{\rm rf}$ in the atom beam holes of the science chamber are the cause of the structure in the interference spectrum, with a signal that is about 3 to 4 orders of magnitude larger than the signal expected from the $\mathcal{E}_{PNC}$ transition alone.  We explore this field component, and suggest a strategy for reducing it, in the following section.

  \begin{figure}
 \begin{centering}
 	  \includegraphics[width=0.45\textwidth]{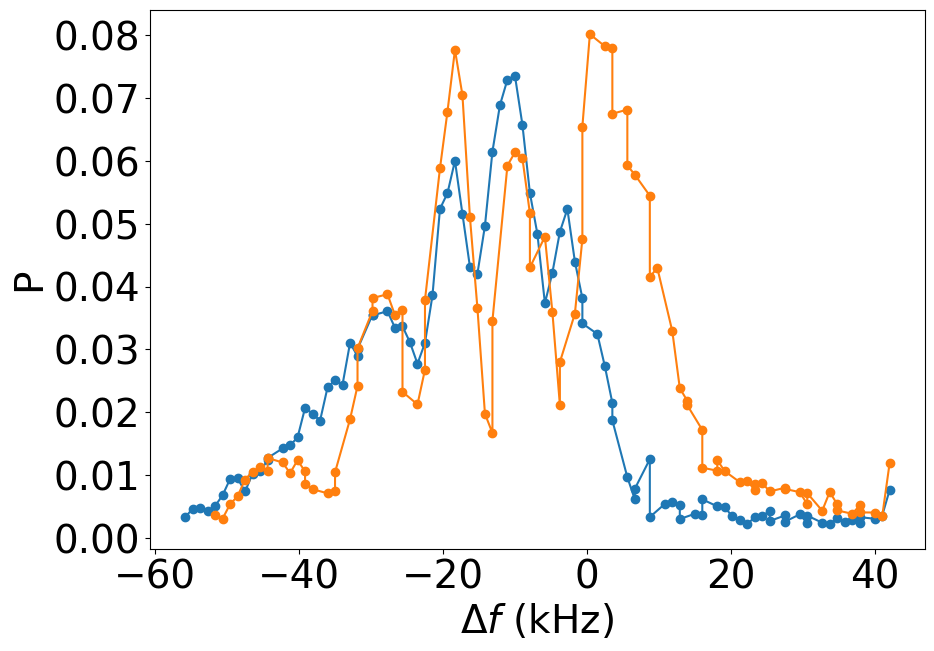}\\

  \caption{The interference amplitude when the two rf inputs are in phase with one another (orange trace) or out-of-phase (blue trace).  Each data point represents the average signal over 2 seconds. }
        \label{fig:interrfb}
 \end{centering}
 \end{figure}

\section{Minimizing $h_z^{\rm rf}$; A four-input cavity design}\label{sec:four_input_cavity}
As indicated by Eq.~(\ref{eq:V_m2}), any rf magnetic field pointing in the $z$-direction would easily produce an M1 signal that could overwhelm the $\mathcal{E}_{PNC}$ signal if not addressed. In an ideal cylindrical cavity operating on a TM mode, of course, $h_z^{\rm rf}$ is identically zero at all positions within the cavity. The cavity is not ideal, however, due to the presence of the atom holes and the rf power coupling channels. We have not identified any field reversal strategy or other technique that might be used to separate this $h_z^{\rm rf}$ contribution from the primary PNC signal of our measurement.  Therefore the magnitude of $h_z^{\rm rf}$ must be reduced. We used numerical simulations to determine the impact of these features on the field pattern of the rf cavity. The results of these numerical models for $h_z^{\rm rf}$ for the entire cavity are shown in Fig.~\ref{fig:comsolall}(b).  

Our primary concern was fields in the vicinity of the entrance and exit holes for the atoms, since the atoms are close to these openings. 
On the cavity axis, the average value of $h_z^{\rm rf}$ is less than $10^{-8}$ A/m, but this increases in magnitude closer to the atom hole walls.  
We show a color plot of the $h_z^{\rm rf}$ field component near the atom holes in Fig.~\ref{fig:Hzentry}.
\begin{figure}
 \begin{centering}

    \includegraphics[width=0.45\textwidth]{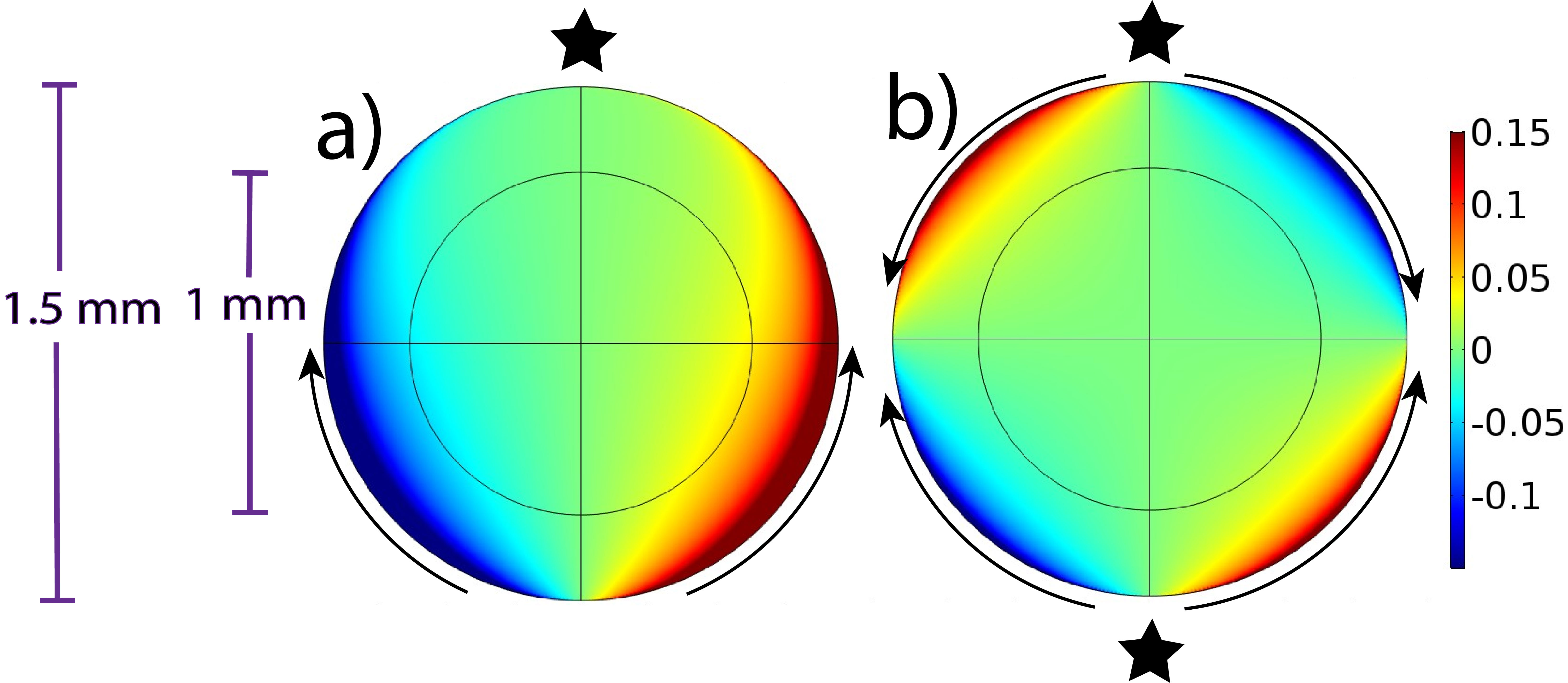}\\

   \caption{ Color plots of $h_z^{\rm rf}$ over the cross section of the atom entry port (1.5 mm diameter) at the surface of the rf cavity, in  units of (A/m). In (a), only the upper excitation chamber is excited, while in (b), both chambers are driven with equal amplitude and phase. The stars denote active excitation ports and the inner black circle is the intended size (1 mm diameter) and location of the atom beam. The atom beam propagates into the page (z-direction). }
         \label{fig:Hzentry}
  \end{centering}
  \end{figure}
Within the 1.0 mm diameter atomic beam region (indicated by the black circle), $h_z^{\rm rf}$ is less than $0.05$ A/m in Fig.~\ref{fig:Hzentry}(a). 
In Fig.~\ref{fig:Hzentry}(a), only the upper excitation chamber is excited.  
The magnetic field in the figure is consistent with an oscillating current flowing upward towards the excitation chamber, and splitting to both sides of the atom hole, as shown by the black curved arrows.
Simulations show that $h_z^{\rm rf}$ increases with larger atom hole size.  

The color map in Fig.~\ref{fig:Hzentry}(b) shows $h_z^{\rm rf}$ when both chambers are excited. The magnitude of $h_z^{\rm rf}$ is reduced by a factor of two due to cancellation between the sources, while $e_z^{\rm rf}$ in the interior of the science chamber increases by a factor of 1.4. Again, this field pattern is consistent with currents from both of the active ports, splitting and wrapping around the atom hole, as shown with black curved arrows in the figure. 
Within the 1 mm diameter ring, representing the atomic beam, the maximum $h_z^{\rm rf}$ with both rf inputs active is 0.02 A/m, reduced from the value with only one input active by a factor of $\sim0.4$.

One metric of the maximum tolerable field amplitude $h_z^{\rm rf}$ in this measurement is the value of $h_z^{\rm rf}$ for which $|V_m|$ (given by Eq.~(\ref{eq:V_m2})) is equal to $|V_{\rm PNC}|$ (given by Eq.~(\ref{eq:V_PNC})). 
From this prescription, we estimate that the ratio $h_z^{\rm rf}/e_z^{\rm rf}$ should be less than $3$ pS.  For $e_z^{\rm rf} = 70$ kV/m, as shown in Fig.~\ref{fig:comsolall}(c), $h_z^{\rm rf}$ should be less than $0.15 \: \mu$A/m by this prescription.
This estimate of the maximum $h_z^{\rm rf}$ may be overly pessimistic, however, in that this field is restricted to regions close to the atom holes, of length $\sim$1 mm, and is of opposite phase at the entrance and exit holes. 
Coherent excitation in two consecutive spatially-separated interactions can result in a Ramsey fringe pattern~\cite{foot2005atomic}, which, in this case, is spectrally broad, with a null at the center of the spectrum. By contrast the spectrum of the $\mathcal{E}_{PNC}$ signal will be $\sim$10 times narrower, with its maximum at line center.  Furthermore, averaging over the cross section of the atomic beam should reduce the magnetic signal significantly.  Still, separating the $\mathcal{E}_{PNC}$ signal from the M1 signal is expected to be challenging. 

In order to better determine the maximum $h_z^{\rm rf}$ for which we can still perform an accurate measurement of $\mathcal{E}_{PNC}$, we calculated the spectrum of the population modulation amplitude $P$ vs.\ the frequency detuning from resonance $\Delta f$, shown in Fig.~\ref{fig:DelPopSpec}.
\begin{figure}
 \begin{centering}
\includegraphics[width=0.45\textwidth]{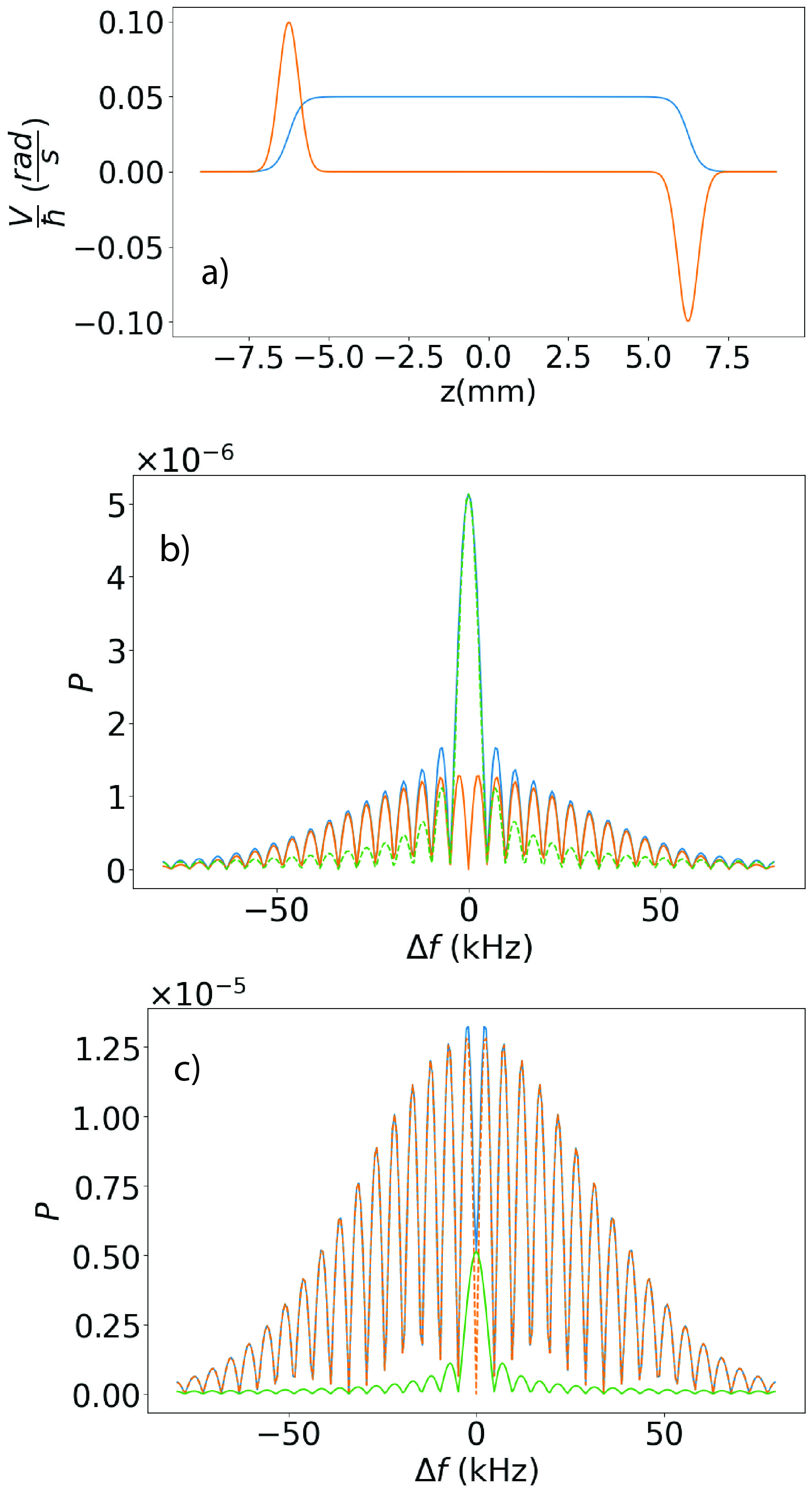}\\
   \caption{(a) The interaction amplitudes $V_m$ and $|V_{\rm PNC}|$ across the rf cavity. (b-c) The amplitude of the population modulation $P$ spectra. In (b), the magnetic dipole interaction $V_m = 2 |V_{\rm PNC}|$, while in (c) $V_m = 20 |V_{\rm PNC}|$. Each figure shows the spectra with the PNC interaction alone (green), the magnetic interaction alone (orange), and with both interactions (blue).  }
         \label{fig:DelPopSpec}
  \end{centering}
  \end{figure}
We determine this spectrum by numerically integrating the  equations of motion of the amplitudes $c_g(t)$ and $c_e(t)$, Eqs.~(\ref{eq:equationofmotionexcited}) and (\ref{eq:equationofmotionground}), including the M1 interaction with $h_z^{\rm rf}$ as the atoms enter and exit the rf cavity, and the $\mathcal{E}_{PNC}$ interaction with $e_z^{\rm rf}$ as the atoms traverse the width of the cavity. $P$ in these plots includes the effect of averaging over the spatial profile of the atom beam, and over the velocity distribution of the atoms. In Fig.~\ref{fig:DelPopSpec}(a), these potentials along the cavity axis are shown, with the maximum value of $|V_m|$, which is significant only near the atom ports, equal to $2|V_{\rm PNC} |$, which is constant across the interior of the rf cavity, but zero elsewhere. (For $\varepsilon_z^{\rm rf} = 70$ kV/m, this condition is met when $h_z^{\rm rf} = 0.3 \ \mu$A/m.) 
 The spectrum of $P$ vs.\ $\Delta f$ of the rf field is shown in Fig.~\ref{fig:DelPopSpec}(b). The different traces in this figure show the population modulation amplitude with the PNC interaction alone (green), the magnetic interaction alone (orange), and with both interactions (blue). Note that the PNC interaction stands out as a narrow, central peak, even though the peak magnitude of the M1 interaction is twice as large as the PNC interaction, while the magnetic interaction produces relatively low-level sidelobes. The spacing of these sidelobes is consistent with Ramsey fringes resulting from the magnetic interaction at the entrance and exit tunnels.  
Note also that the magnitude $P$ at line center is the same with or without the $V_m$ interaction present.  At zero detuning, the Ramsey fringe pattern of the $V_m$ interaction has a null (since $V_m$ at the two atom tunnels are equal in magnitude but opposite in sign), and $V_m$ and $V_{\rm PNC}$ are $\pi/2$ out of phase with one another (since $\mathcal{E}_{\rm PNC}$ is imaginary). When $V_m$ is increased by a factor of ten ($|V_m| \sim 20 |V_{\rm PNC}|$), the side lobes increase in magnitude, but the value of $P$ at line center is still unchanged by the presence of the $V_m$ interaction. (See Fig.~\ref{fig:DelPopSpec}(b).)  While this argument holds when $h_z^{\rm rf}$ is of equal amplitude at the two atom holes, the signal due to $V_m$ does not vanish if $h_z^{\rm rf}$ differs at the two atom holes. We do not explore that in any detail here, but it does emphasize the point that $h_z^{\rm rf}$ should be reduced to the extent possible.

The analysis of the last paragraph suggests that a successful measurement of $\mathcal{E}_{\rm PNC}$ is possible if $V_m$ can be reduced to $\sim10 |V_{\rm PNC}|$.  With the current 2-port rf cavity, however, the maximum value of $h_z^{\rm rf}$ within the 1 mm diameter circle of the atom beam appears to be $\sim$0.01 A/m, which leads to $V_m \sim 6 \times 10^4 |V_{\rm PNC}|$.  We therefore have been exploring methods by which we can reduce the magnetic field $h_z^{\rm rf}$ near the atom holes.  One promising avenue is to increase the number of rf ports that couple power into the rf cavity. This change is suggested by the one feature of the cylindrical cavity that is not cylindrically symmetric; the rf power coupling ports. As stated earlier, the wall currents on the interior of the perfect cavity all point radially outward from the atom holes.  The presence of $h_z^{\rm rf}$ is due to non-radial currents, as shown in Fig.~\ref{fig:Hzentry}. With the addition of two more rf ports, for a total of four, equally spaced about the cavity, the currents strongly cancel one another, and the $h_z^{\rm rf}$ is significantly reduced. We show this cavity design in Fig.~\ref{fig:interrf}. 
\begin{figure}
 \begin{centering}
 	  \includegraphics[width=0.25\textwidth]{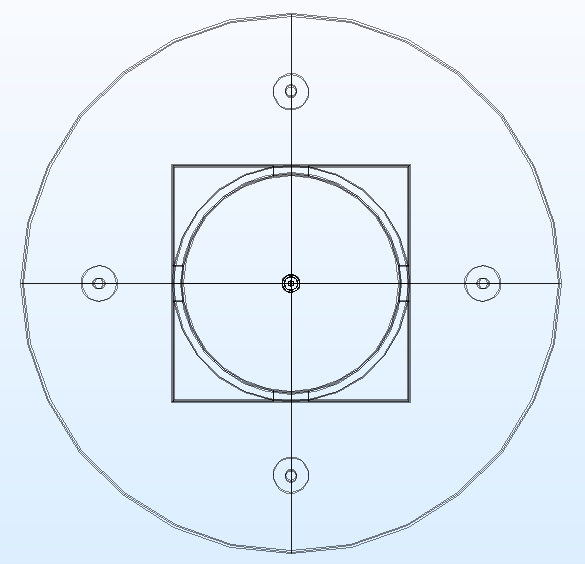}\\

  \caption{ Proposed 4 port cavity design. The central circular outline is the science chamber, about which are four equally-spaced rf input ports. Power is coupled from the outer excitation chamber into the central cavity via four coupling channels.
      \label{fig:interrf}}
 \end{centering}
 \end{figure}
The science chamber is the round cavity in the center, surrounded by the excitation chamber, which is excited by four rf inputs.  Coupling channels, positioned between each of the rf inputs and the central axis, connect the excitation chamber with the science chamber. Numerical simulations show that this reduction of $h_z^{\rm rf}$ is more than a factor of $10^2$ on the edge of the atom beam,
while increasing the electric field $e_z^{\rm rf}$ by a factor of 3. For the four rf inputs the ratio $h_z^{\rm rf} / e_z^{\rm rf}$ is less than 40 pS and $|V_m| \lesssim 20 |V_{\rm PNC}|$ on the center axis of the cavity (max $h_z=10 \mu $A).  A color map of $h_z^{\rm rf}$ near the atom hole for the 4-input cavity is shown in Fig.~\ref{fig:Hzentrt4port}.  
 \begin{figure}
 \begin{centering}
   \includegraphics[width=0.45\textwidth]{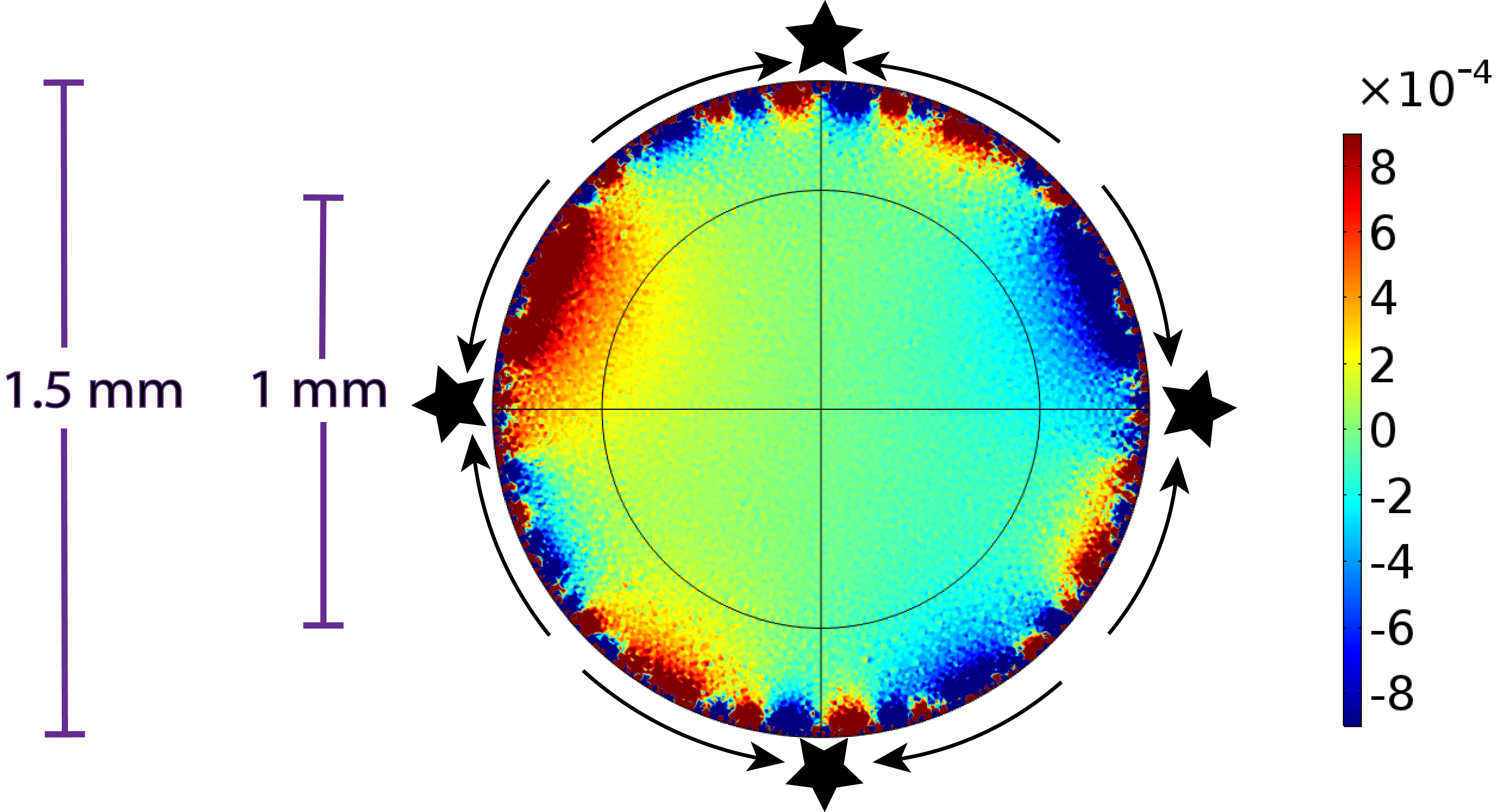} \\

    \caption{  Color plot of $h_z^{\rm rf}$ (in units of A/m) over the cross section of the atom entry port (1.5mm) for the newly designed 4-port rf cavity. 
    The stars denote excitation ports and the inner black circle is the intended size (1 mm diameter) and location of the atom beam. The atom beam propagates into the page (z-direction). 
  \label{fig:Hzentrt4port}}
\end{centering}
\end{figure}
For the following reasons, we believe that the magnetic fields are actually smaller than computed.
(1) As shown in Fig.~\ref{fig:Hzentrt4port}, the field results are quite noisy, they show strong variation over distances much less than the wavelength of the field, and they do not display the four-fold symmetry of the cavity, unlike the results in Fig.~\ref{fig:Hzentry}. (2) These are extremely low-magnetic field amplitudes, $\sim 10^{-4}$ A/m, while the electric field along the central axis is $2 \times 10^5$ V/m. (3) The results are sensitive to details of the grid pattern for the numerical computation, and the $h_z^{\rm rf}$ values have decreased with each decrease in the grid spacing. 
(A further decrease is not feasible with the constraints of available computer memory.) For these reasons, we expect that these noisy $h_z^{\rm rf}$ results are likely an overestimate of the magnetic fields within the 1 mm diameter atom beam, and the actual fields are even smaller.  

We therefore conclude that the extra-cavity fields are sufficiently small and that they do not affect the signal that we have obtained so far.

\section{Conclusion}\label{sec:Conclusion}
We have successfully observed the coherence of the rf fields in the rf cavity with a pair of Raman lasers. 
This measurement makes use of the pure rf fields of the TM$_{010}$ mode of a cylindrical cavity. The critical features of this mode are its high-amplitude, uniform electric field and small magnetic field $h_z^{\rm rf}$ along the axis.  With implementation of the four-input cylindrical cavity described in Sec.~\ref{sec:four_input_cavity}, we expect further reduction of this field component to the level necessary for a successful measurement of $\mathcal{E}_{PNC}$ due to the anapole moment.  We also plan improvements in the uniformity of the transverse magnetic fields $B_x$ and $B_y$. Using current measurements of the noise level ($\sim$21 $\mu$V$/ \sqrt{\rm Hz}$) and the expected signal size (3.5$\mu$V), we project a 10\% measurement can be reached with $\sim$3600 seconds of integration time.

We would like to thank Kevin Webb for help with the numerical simulations and Jason McKinney for insightful conversations about rf cavity design.
This material is based upon work supported by the National Science Foundation under Grant Number PHY-1912519. 

\bibliography{biblio}

\end{document}